\documentclass[conference,compsoc]{IEEEtran}
\ifCLASSOPTIONcompsoc
  \usepackage[nocompress]{cite}
\else
  \usepackage{cite}
\fi
\ifCLASSINFOpdf
\else
\fi
\usepackage{pgfplots}
\pgfplotsset{compat=1.18} 

\usepackage{enumitem}

\usepackage{adjustbox}
\usepackage{float}
\usepackage{multirow}
\usepackage[section]{placeins}
\usepackage{pgf-pie}  
\usepackage{array}
\usepackage{xurl}
\usepackage{subcaption}
\usepackage{amssymb}
\usepackage{tikz}

\usetikzlibrary{shapes,arrows.meta,chains}

\definecolor{ibmyellow}{HTML}{ffb000}
\definecolor{ibmorange}{HTML}{fe6100}
\definecolor{ibmmagenta}{HTML}{dc267f}
\definecolor{ibmindigo}{HTML}{785ef0}
\definecolor{ibmblue}{HTML}{648fff}

\usepackage{hyperref}
\hypersetup{
    colorlinks=true,
    linkcolor=ibmmagenta,
    filecolor=ibmorange,      
    urlcolor=ibmblue,
    citecolor=ibmmagenta,
    }

\usetikzlibrary{cd}

\begin{document}
%
\title{Sark: Oblivious Integrity Without Global State}

\author{\IEEEauthorblockN{Alex Lynham}
\IEEEauthorblockA{UCL Computer Science\\
University College London \\
66-72 Gower St\\ 
London \\
WC1E 6BT\\ 
United Kingdom}
\and
\IEEEauthorblockN{David Alesch}
\IEEEauthorblockA{UCL Computer Science\\
University College London \\
66-72 Gower St\\ 
London \\
WC1E 6BT\\ 
United Kingdom}
\and
\IEEEauthorblockN{Ziyi Li}
\IEEEauthorblockA{UCL Computer Science\\
University College London \\
66-72 Gower St\\ 
London \\
WC1E 6BT\\ 
United Kingdom}
\and
\IEEEauthorblockN{Geoffrey Goodell}
\IEEEauthorblockA{UCL Computer Science\\
University College London \\
66-72 Gower St\\ 
London \\
WC1E 6BT\\ 
United Kingdom}}


%


\maketitle

\begin{abstract}
In this paper, we introduce \textit{Sark}, a reference architecture for transferring unforgeable, stateful, oblivious (USO) assets. We describe the motivation, design, and implementation of the core subsystems of Sark, \textit{Porters}, which accumulate and roll-up commitments from \textit{Clients}, and \textit{Sloop}, a permissioned, crash fault-tolerant (CFT) blockchain system. We analyse the operation of the system using STRIDE threat analysis, and the `CIA Triad': Confidentiality, Availability, and Integrity. We then introduce the concept of \textit{local centrality} and address design trade-offs related to decentralization. Finally, we suggest future work on Byzantine fault-tolerance (BFT), and mitigating the local centrality of Porters.
\end{abstract}


%
\IEEEpeerreviewmaketitle

\section{Introduction}
Sark is a distributed system that offers oblivious, non-custodial management of assets with remote integrity.\footnote{We use the NIST definition of \textit{integrity}, ``guarding against improper information modification or destruction and ensuring information non-repudiation and authenticity.''~\cite{nist_cia} `Integrity' maps neatly on to a discussion of immutability, both in the strict sense~\cite{isoblockchain} and the sense of \textit{practical immutability}, wherein data can be changed subject to the governance demands of a given network's governance topology.~\cite{lynham_goodell_immutability_springer}} It has many of the benefits of common blockchain stacks, such as finality that serves as proof against opportunistic rewrite, while meaning a different set of limitations, such as reduced data availability---only a single relay (Porter) can handle reads and writes for a USO asset\footnote{Unforgeable, stateful, oblivious (USO) assets are self-contained; their properties are described in Section 3.1. Obliviousness serves as both a way of ensuring privacy, but also transaction efficiency---all parties are not required to know all states. Where `privacy' (after Nissenbaum ~\cite{nissenbaum}) is a function of social norms and transmission context, `obliviousness' is the \textit{requirement} for a subsystem or agent to be oblivious.} that is tethered to it for its integrity.\footnote{This results in what we call \textit{local centrality}---an asset having dependence on a single relay, or Porter, specified \textit{in} the asset, before transfer.} In return, Sark is oblivious by default, which we argue makes it systemically more robust at the governance topology level by supporting agent independence.\footnote{The \textit{governance topology} is ``the structure of decision making power in the network''~\cite{KOINE:LynGoo2025} and serves as a counterpoint to the physical and technical structure, or \textit{network topology}.}


The key contribution this paper makes is:

\vspace{0.3em}
\textit{Describing the design and architecture of Sark, a local-first, oblivious integrity system that generates proofs that can be verified with a blockchain.}\footnote{We define a \textit{local-first system} as a system where finality from the perspective of a user is principally dependent on a single node in the system's \textit{network topology} (physical structure), to which the user is directly connected---presumably via an API or client software. Thus typical blockchain systems might be described this way if say, run in development mode as a single node on a local machine, but not in production where a consensus process must occur.}
\vspace{0.3em}

Section 2.1 provides an overview of Sark's motivation and high-level design objectives, and 2.2 situates it within a discourse of oblivious payments, decentralization and immutability. Sark's design addreses the observation that the \textit{network topology} and \textit{governance topology} of most existing public, permissionless blockchain networks introduce inefficient transaction cost economics through externalities. Requiring governance to account for these externalities results in `decentralization theatre,'~\cite{vergne} since performative decentralization is cheaper\footnote{Both in terms of literal contracts or agreements with counterparties, and the cost of co-ordination, which increases as authority disperses. If a Foundation keeps tighter, more centralized control of a network, or if Validators form informal guilds,~\cite{lynham_goodell_immutability_springer} then this cost of co-ordination is lower.} than actual dispersion of authority.~\cite{KOINE:LynGoo2025} Sark reduces governance overhead for the user by pushing functions that deliver settlement finality to the system edge and requiring greater trust in a smaller number of system nodes. Obliviousness is also a hard check on certain types of opportunistic adversarial actions.

In Section 3 we describe the architecture, protocol, and implementation of Sark, offering a summary of its design goals.\footnote{Initially, we analyse Sark in the context of being run in a permissioned or consortium context. This is discussed in Sections 4.1-4.3.} In Section 4 we critically analyse our work, describing a threat model in Section 4.1, before comparing Sark's design with existing systems on three axes: \textit{Confidentiality}, \textit{Availability}, and \textit{Integrity}, in Section 4.2.\footnote{The so-called `CIA Triad' of data security.~\cite{nist_cia}} We then analyse global centrality and local centrality in Section 4.3.\footnote{Global centrality being dependence on a single node (or small number of nodes) at the centre of the network, and local centrality being dependence on a single node (or small number of nodes) at the edge.} Finally, in Section 4.4 we describe `communities' of Porters.

In concluding, we note that although Sark has local centrality, its lack of global state makes it a radically more decentralized system in terms of its \textit{governance topology} than those that depend on a global ledger. Sark has virtually no dependency on global governance, and Porters (relays) are independent, not requiring co-ordination or consensus to operate. Whether or not users find obliviousness to be a source of trust and confidence remains to be seen. This difference in institutional and trust topology is implied by the work of Davidson et al.~\cite{DAVIDSON_DE_FILIPPI_POTTS_2018} and De Filippi et al.~\cite{DEFILIPPI2020101284} (amongst others) on the blockchain as institution.

\section{Context}

\subsection{Motivation}

We hypothesise that existing systems that require a global state machine create externalities in the form of costs that must be either systematically recovered or continually subsidised. In the case of a public, permissionless blockchain, these costs arise from the network topology and, we hypothesise, are expected by network participants to be recovered in the governance topology by network governance, introducing a coordination problem that might not scale. In the case of a private or consortium ledger, the existence of these costs can impact the economic relationships among consortium members.

However, a ledger with public visibility introduces externalities related to the visibility of transactions even if the transactions it contains are private by design, motivating a design that is not just private, but \textit{oblivious}, in which the ledger operator does not learn what is being transacted or the parties involved. Not only does such an approach substantially remove the externalities associated with public visibility, but it also means that opportunistic actions in the network topology are checked by removing the potential to incentivise, quantify, or verify the performance of opportunistic adversarial actions. Removing such potential actions reduces the cost of coordination at the governance level, since there is less scope for misbehaviour.

The design that serves as the motivation for Sark is a regulation-compatible oblivious asset system that preserves user confidentiality,\footnote{``Preserving authorized restrictions on information access and disclosure, including means for protecting personal privacy and proprietary information,'' per the NIST definition.~\cite{nist_cia}} and can be used flexibly, not only in the context of digital currency, but also to issue other non-monetary digital assets, such as diplomas or identity proofs.~\cite{goodell_et_al_uso} Indeed, any existing blockchain use-case that requires or employs NFTs in its design or implementation can potentially be addressed by this system, providing that it does not require public visibility. Sark\footnote{A mirror of the repository is open-source and available at \url{https://forge.cs.ucl.ac.uk/Sark/Sark}.} is a reference implementation that splits the architecture into three key subsystems, with the intention that each individual subsystem could be replaced if necessary, either to address a different problem space or to alter the trade-offs in the system. For instance, an operator may wish to replace Sloop, the permissioned CFT blockchain component, with a permissioned BFT blockchain, or even a permissionless ledger in some circumstances.

\subsection{Related work}

In 1983, David Chaum proposed a privacy-preserving payments system,~\cite{chaum_1983} the beginning of a family tree of academic and practical work that informs the design and implementation of Sark. Working papers involving the Swiss National Bank~\cite{chaum2021issuecentralbankdigital} and the adoption of GNU Taler have since validated this general approach to asset design, while in the permissionless space, oblivious blockchain-based systems that rely on a local-first approach such as Penumbra exist,~\cite{penumbra_docs} as well as more widely-known zero-knowledge blockchains such as Zcash and Monero.

More recently, Zero-knowledge rollups have become commonplace as a scaling `layer 2' (L2)\footnote{``A layer 2 is any off-chain network, system, or technology built on top of a blockchain to help extend its capabilities.''~\cite{chainlink_l2}} solution for congested or expensive `layer 1' (L1) blockchain systems, such as Ethereum. Prominent examples of L2s include Polygon zkEVM, dYdX v3, Linea and Starknet. These systems are somewhat comparable to the Sark architecture, although their primary goal is aggregation, and they rely on the continued existence of a single, specific chain (in this case, Ethereum) to guarantee the correctness of their transactions.

However, many of these L2 systems are implemented as blockchains themselves, with their own consensus protocols. This is different to Sark, where Porters (the integrity-providing relay nodes that add asset updates to their state tries) do not \textit{require} a consensus process, since USO assets are self-contained and maintain their own state. Porters do contain an internal ledger system that creates Merkle roots, but Porters are independent nodes and do not depend on each other to secure the integrity of assets tethered to them. 

A better analogy might be to Proposer Block Separation in Ethereum, with Porters effectively separating `block' creation from the consensus layer. However, an additional difference comes at the asset layer---USO assets are stateful, and are not tied to the implementation of any one ledger.

Sark anticipates the use of an external blockchain system to provide an external Proof of Inclusion (POI) for stronger protection against double spending or to ``saturate'' or complete the USO asset's Proof of Provenance (POP). Once an update has been registered by a Porter relay,\footnote{Using the terminology of Tendermint, on block finalization, the \texttt{CommitBlock} message emitted by the consensus client would be used by the Porter to memorialise the ledger root in which its own transaction trie root (or TTR, see Section 3) was stored.} the external ledger might ``go away'' (e.g. halt or fork), and the asset could still be validated locally, without needing to check the external ledger.  Only a valid, signed Merkle root from a ledger that includes the Porter's \textit{transaction trie root} (TTR) is required to prove inclusion.\footnote{This has the benefit that with appropriate middleware, as described in Section 3, Porters are compatible with many existing ledger designs.} The primary trust relationship is between the user and the Porter operator, not the whole network; instead, the role of the network is to strengthen that trust.

\subsubsection{Decentralization}
Although there are many models for decentralization, including the Nakamoto Coefficient\footnote{Defined as, ``the minimum number of entities in a given subsystem required to get to 51\% of the total capacity\ldots [or] the operative threshold''~\cite{srinivasan_lee} for a given subsystem.} and Edinburgh Decentralization Index (EDI),~\cite{edi}\footnote{The EDI is a quantitative stratified model, whereas their MDT (Minimum Decentralization Test) is more directly comparable to the Nakamoto Coefficient. Its definition is looser, ``A blockchain system fails the Minimum Decentralization Test (MDT) if and only if there exists a layer for which there is a single legal person that controls a sufficient number of relevant parties so that it is able to violate a property of interest.''~\cite{edi_ovezik2024sokstratifiedapproachblockchain} This means it can cover coercion and collusion in a way that is complementary with the agent dynamics described in qualitative studies.~\cite{KOINE:LynGoo2025}} we use the two-axis model advanced by Lynham and Goodell. This is a model for decentralization in systems that include a blockchain, which operates on two axes:

\begin{itemize}
    \item \textbf{Network topology}, the physical and technical structure of the network.
    \item \textbf{Governance topology}, the structure of decision making power in the network.~\cite{KOINE:LynGoo2025}
\end{itemize}

Though originally applied to decentralization, this model can be used to analyse integrity in distributed systems.\footnote{One might consider the heuristic that one is the protocol, and the other is the governance of the protocol.} Although there is no formal relationship between decentralization and immutability, both can be analysed using it.\footnote{Where decentralization is the output of dispersion of authority on the variables that make up the model's two topologies, and immutability is the output of the two topologies, as expressed on ledger state.}

\subsubsection{Immutability, Governance and Trust}
Lynham and Goodell argued immutability was less robust than the absolute immutability of the ISO definition (ISO 22739:2024 3.51); the ``property of a distributed ledger (3.23) wherein ledger records (3.55) cannot be modified or removed once added to that distributed ledger.''~\cite{isoblockchain} and, based on analysis of interviews, proposed characterizing the immutability found on these systems in terms of \textit{practical immutability}, ``data [are] immutable, except where the data [are] deemed to be illegitimate,'' visualising ledger governance thus:~\cite{lynham_goodell_immutability_springer}

\smallskip

\begin{center}
\begin{adjustbox}{width={0.9\columnwidth},totalheight={\textheight},keepaspectratio}
\begin{tikzcd}[column sep=5em]
    Agent \arrow{r}[]{Trust} & {Governance} \arrow{r}[]{Legitimacy} & {Ledger\:State}
\end{tikzcd}
\end{adjustbox}
\end{center}

\smallskip

In such systems, \textit{trust} and \textit{confidence} are key to their operation. According to Williamson,~\cite{williamson1993} Trust is an agent applying behaviour-cost economics to a transaction. It can manifest calculatively (as bounded rationality and opportunism) or non-calculatively (via interpersonal, structural, or institutional trust).\footnote{Agents may also be irrational, resulting in non-calculative trust.} Confidence (Earle via De Filippi et al.~\cite{DEFILIPPI2020101284}) is a state, a feeling in agents based on a calculative assessment of how a system will behave now, and into the future based on past states, experience, and evidence.

For the average public, permissionless blockchain, the level of calculative and non-calculative trust required for the system to operate is high, while at the same time it is often only possible to quantitatively state that a network is \textit{not in immediate danger of takeover} on the basis of its topology. Many security mechanisms, such as slashing in (Delegated) Proof-of-Stake\footnote{For a definition of Proof-of-Stake and Delegated Proof-of-Stake, see Bashir.~\cite{Bashir2022} Both weight validators according to bonded stake, but Delegated Proof-of-Stake allows users (stakers) to bond their stake permissionlessly to validators. This is the default in Cosmos SDK chains.} are only applied ex-post, and cannot definitively secure finality.~\cite{KOINE:LynGoo2025} The result of this is `decentralization theatre',~\cite{vergne} which is essentially performative, since it makes ex-ante claims about the operation of the system in a crisis. Moreover, in light of high-profile rewrite events it seems likely that many agents lend their trust to a blockchain's governance topology and deem its ledger state as legitimate only if it can be rewritten by those same governance processes. This is perhaps why fieldwork has identified emergent, self-organised regulation and professionalization among blockchain node operators~\cite{lynham_goodell_immutability_springer} of the kind described by Williamson.~\cite{williamson1993}

A valid question, then, is: \textit{What is the purpose of decentralization?}  If the goal is systemic robustness, then that could be better expressed as reducing the opportunity for governance overreach,\footnote{For example, in high-profile governance-mandated rewrite events such as the Ethereum DAO hack and subsequent hard fork, or Juno's Proposal 16 token seizure in the Cosmos Ecosystem.~\cite{lynham_goodell_immutability_springer}} or ensuring agent independence.

The design of Sark seeks to avoid these pitfalls by not requiring globally coordinated state. This means that not only the locus of integrity, but also the locus of governance that a user faces has less complexity. Obliviousness in the design of the transaction protocol further reduces the externalities associated with governance, facilitating more efficient transaction cost economics. This is because obliviousness reduces the required trust threshold and helps to ensure agent independence. Compared to blockchain systems that process and validate each transaction, oblivious systems incur strictly lower costs in terms of network operation when validators are added. Simply put, a user does not need to trust these additional agents in Sark, whereas in a public permissionless blockchain network, they do.

\section{Design, Implementation and Architecture}

The design goals of the Sark system are closely tied to the USO asset itself. In a sense, the system exists solely to support the deployment of a USO asset. Its goals are:

\begin{enumerate}
    \item Support unique assets with cash-like properties.\footnote{Such as fungibility, confidentiality, and user self-custodianship without requiring an intermediary.}
    \item Protect against double spending.
    \item Enable assets to have self-contained, independent state that an owner can verify without external services.
    \item Substantially limit the need for global coordination in the \textit{network topology}.
    \item Deliver ``obliviousness by design'' in addition to ``privacy by design.''~\cite{cavoukianprivacy}
\end{enumerate}

The USO system can also be used to address adjacent problems, such as credentialing, identity proofing, and implementing transfers for Real World Assets (RWAs), such as tokenized securities, where an audit trail is required.

\begin{figure}[t!]
    \centering
    \includegraphics[width=0.95\columnwidth]{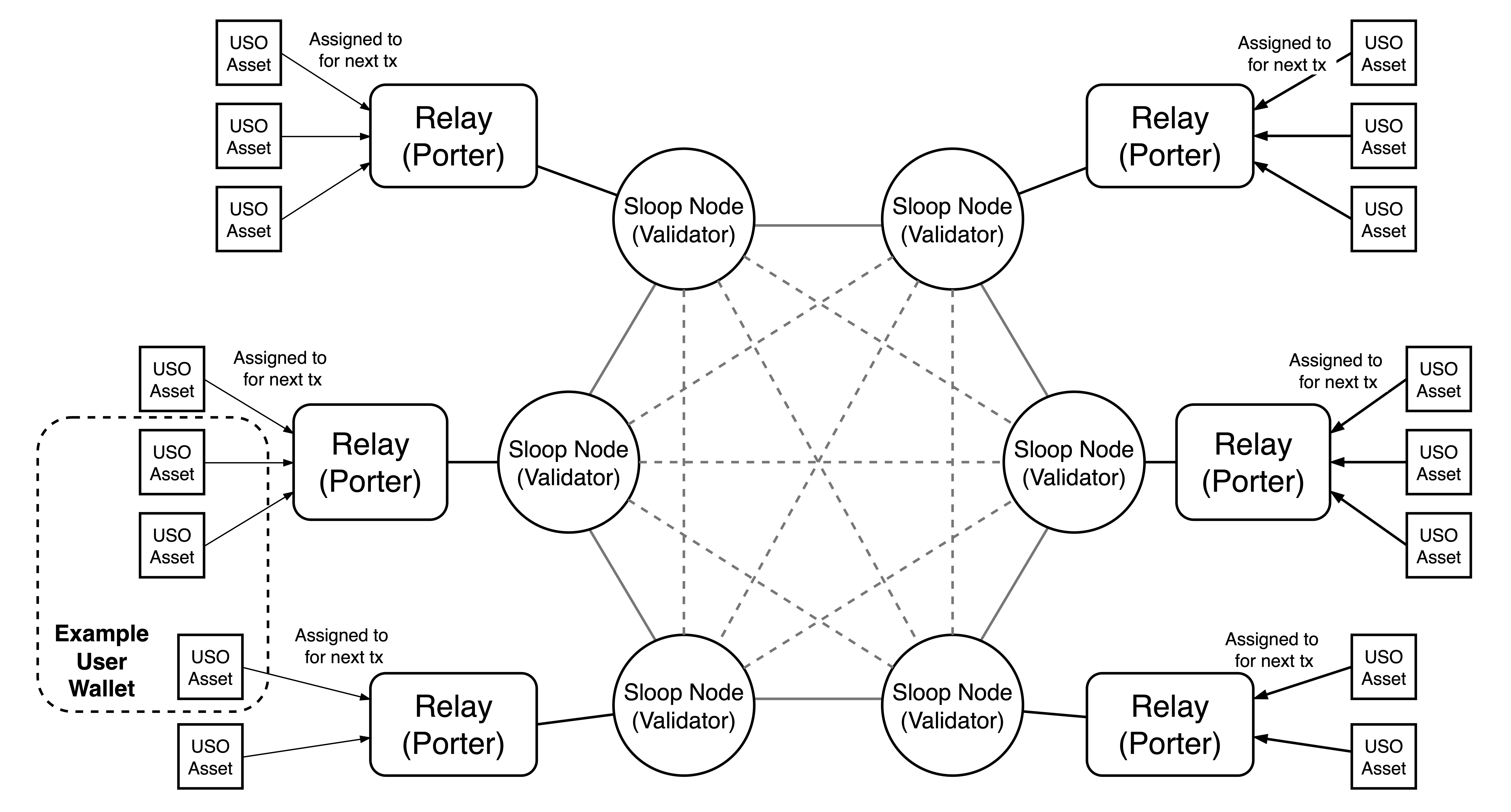}
    \caption{Sark Architecture}
    \label{fig:sark_arch}
\end{figure}

A high-level overview of the Sark system demonstrates that it has a comparable stack to many other blockchain systems, with Porters performing a similar role to execution clients, and Validators handling consensus in a similar manner to consensus clients (Figure \ref{fig:sark_arch}).\footnote{For example, both the default Cosmos stack~\cite{whatiscosmos} and Ethereum stack~\cite{ethereumnodesandclients} have a division between execution and consensus client.} Sloop, the blockchain subsystem, which comprises a Validator and Ledger, is based on Raft, making it crash fault-tolerant (CFT) by design, not Byzantine fault-tolerant (BFT), though it is possible to replace this component with a BFT consensus client, such as one built on Tendermint.\footnote{The main Tendermint implementation has rebranded to CometBFT, but we refer to the protocol and project as Tendermint to avoid confusion. For a discussion of Sloop's finality properties, see Section 3.5.}

\subsection{USO Assets}
\label{ss:uso}
USO assets are the transactable unit. They have three key properties (the protocol is described in Section 3.2):

\begin{enumerate}

    \item \textbf{Unforgeability}. ``Every asset must be unique, and it can only be created once. No set of adversarial actors can repeat the process of creating an asset that has already been created\ldots This property is required for durability, custodial choice, the choice to have no custodian, local transactions, and time-shifted offline transactions.''~\cite{goodell_et_al_uso}

    \item \textbf{Statefulness}. Every asset has independent state; ``as the state of an asset changes over time, the asset remains unique and unforgeable. No set of adversarial actors, including non-issuer owners, can create a second version of the asset with a different state.''~\cite{goodell_et_al_uso}

    \item \textbf{Obliviousness}. ``Once finality is achieved following the transfer of an asset to a new owner all of the previous owners, including the issuer, have no obligation to know any aspect of its future state changes and transfers. There is no residual risk to the new owner that the transaction will be undone by either a previous owner or the system itself. Note that encryption does not suffice: there must be no requirement to inform previous owners that state changes have occurred, and previous owners must not be required to do any extra work to accommodate those changes. Otherwise, the self-determination and efficient lifecycle requirements would be compromised.''~\cite{goodell_et_al_uso}

\end{enumerate}

\begin{figure}[t!]
    \centering
    \includegraphics[width=0.8\columnwidth]{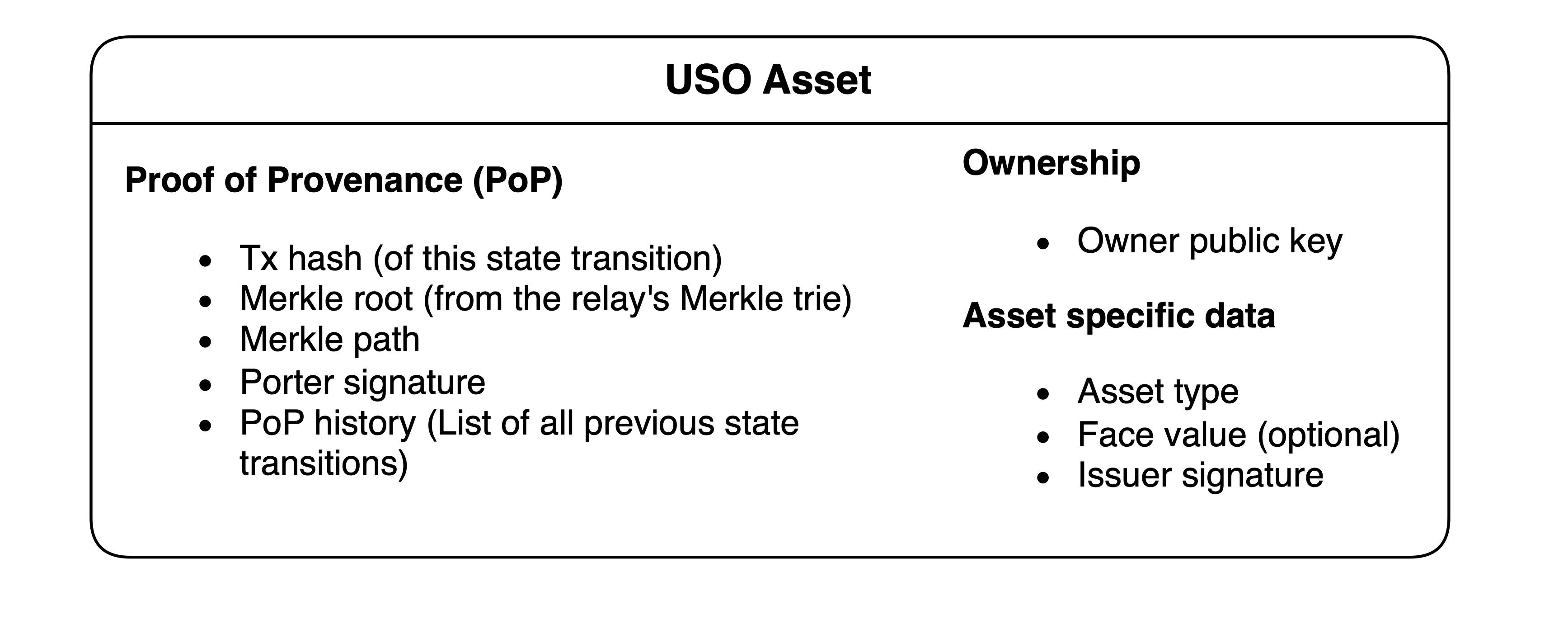}
    \caption{Sark USO Asset}
    \label{fig:uso}
\end{figure}

Unlike tokens recorded on a shared ledger, USO assets maintain internal state and (optionally) metadata about their transactions (Figure \ref{fig:uso}). Each USO is cryptographically unique, unforgeable, and self-contained. Its state evolves through transactions, which are authorised by the owner and processed \textit{only} by the designated relay (Porter), which is specified in advance.\footnote{Since the last asset update (or its initialization) requires it to be tethered to a Porter, duplicate transactions are not possible---the Porter that must include any update in its state is already known and its identity referenced. Thus, the chief threat to an asset holder is Porter equivocation, or downtime. This motivates both including TTRs in an external ledger and minimally-decentralized 'communities' of Porters, described in Section 4.4.}

For each transaction, the relay returns a proof to the asset holder or recipient attesting that the most recent Merkle trie (which we term the \textit{transaction trie}) that it built linked the one-time public key of the owner at the time of the transaction to the specified transaction hash. 

The Porter (relay) can be queried to prove that no fully built Merkle tries since a specified point in its history contain an entry linked to that one-time public key.\footnote{e.g. since the most recent transaction, or issuance of the asset.}

Privacy-by-design is achieved at asset issuance. New USO assets are minted using blind signatures, which prevent any minting authority from linking later ownership or transfers back to issuance and, by extension, the owner. Ordinary transfers do not allow any parties, even acting in collusion, to reveal the identity of the first holder of the asset. Unlike in account-based systems,\footnote{That is, systems that implicitly have an intermediary.} this design ensures that no central party can identify the first holder of the asset.\footnote{Which in an e-cash scenario would be the consumer.}

The protocol that governs issuing and updating USO assets is covered in the next section.

\subsection{The USO Protocol}
To begin creating an asset the initial owner (e.g. a commercial bank in the context of a CBDC) creates the initial update vector $F_0$ containing an arbitrary message ${u_0}$, a reference ${G_{L,i}}$ to a specific root $i$ of an oblivious ledger $L$ and relay (Porter) $G$, and the public key $k_1$ matching a new, one-time private key $k^*_1$, as follows:
\begin{figure}[t!]
    \centering
    \sf\scalebox{0.85}{\begin{tikzpicture}[>=latex, node distance=3cm, font={\sf \small}, auto]
    \tikzset{>={Latex[width=2mm,length=2mm]}}
    \node (n1) at (0,0) {Alice};
    \node (n2) at (4,0) {Provider of $G$};
    \node (n3) at (-4,0) {Bob};
    \draw (n1) -- (0,-3.1) {};
    \draw (n2) -- (4,-3.1) {};
    \draw (n3) -- (-4,-3.1) {};
    \draw[->] (-4,-0.7) -- node[above] {
        $k_{j+1}$
    } (0,-0.7);
    \draw[->] (0,-1.2) -- node[above] {
        $k_j,s(h(F_j),k_j)$
    } (4,-1.2);
    \draw[->] (4,-1.9) -- node[above] {
        $p(G_{L,i},k_j,h(F_j))$
    } (0,-1.9);
    \draw[->] (0,-2.4) -- node[above] {
        $F_j,\pi_j$
    } (-4,-2.4);
    \end{tikzpicture}}
    \caption{Alice registers an update, giving Bob control first and possession later.}
    \label{fig:pop_update}
\end{figure}

\begin{equation}
F_0 \leftarrow (u_0, G_{L,i}, k_1)
\end{equation}

This vector serves as the starting point for the creation of an asset. With it, the initial owner can create an asset $A_0$ by signing it with a long-term key $k_0$ held by some issuer: 

\begin{equation}
A_0 \leftarrow (F_0, s(h(F_0),k_0))
\end{equation}

\begin{figure}[t!]
    \centering
    \includegraphics[width=0.9\columnwidth]{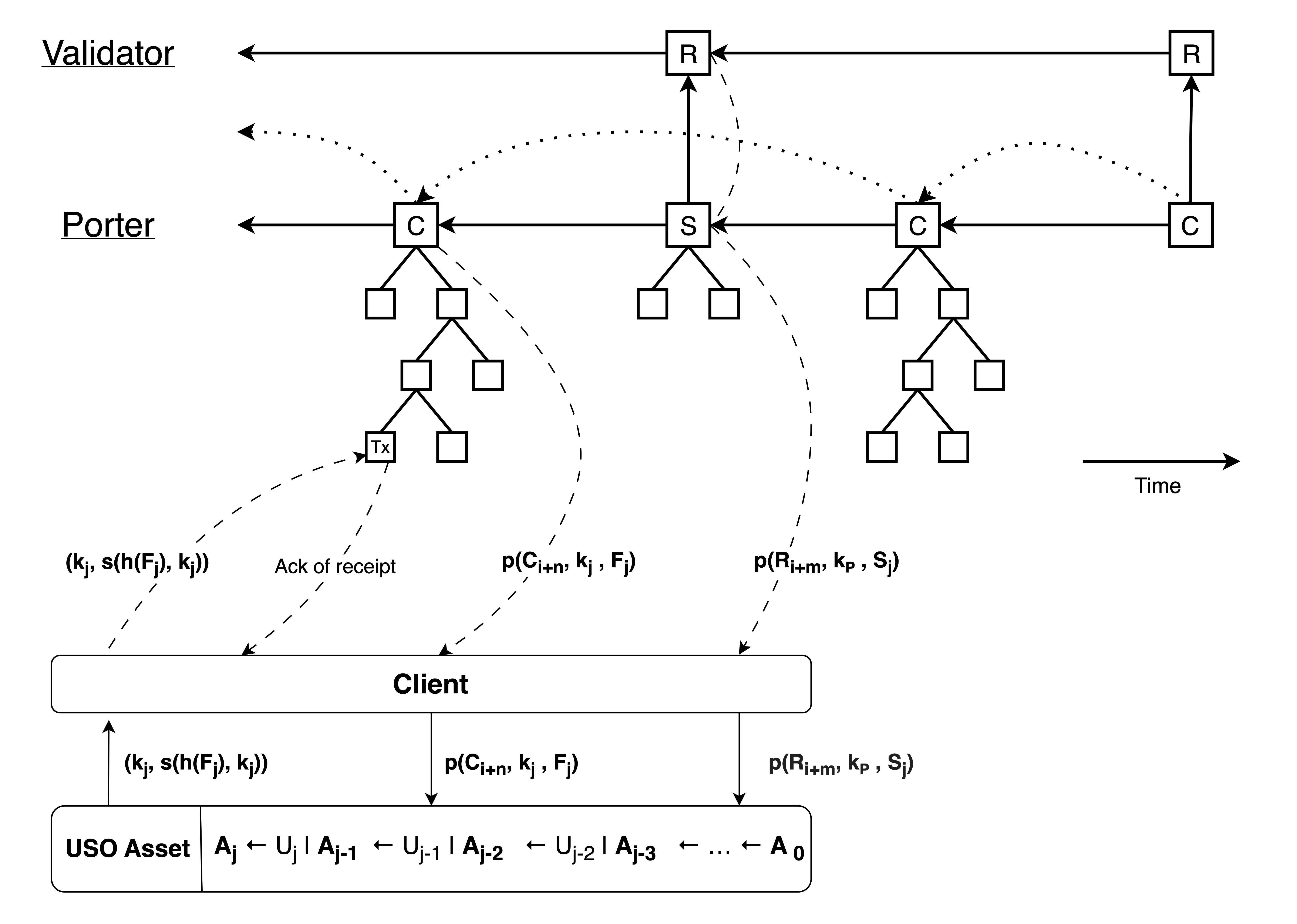}
    \caption{Sark Protocol Diagram}
    \caption*{\tiny Note: C represents completed Merkle trees at the Porter level, S represents snapshots of Merkle trees in production, which can be submitted at any given rate, and R represents the roots at the Validator level.}
    \label{fig:sark_protocol}
\end{figure}
Updating (i.e. transferring) the asset then follows the same schema (Figure \ref{fig:pop_update}). The owner at sequence number $j$ must create an \textit{update vector} $F_j$ containing the same three fields as in $F_0$ and a hash of $A_{j-1}$ and sign it with $k_j$ to create the update $U_j$. To complete the transfer, $k_{j+1}$ in the update $F_j$ \underline{must} be provided by the new owner, as $k_{j+1}$ determines future control. The updated asset $A_j$ is then defined by concatenating the previous version of the asset $A_{j-1}$ with the update:

\begin{equation}
A_j \leftarrow A_{j-1} \vert\vert U_j
\end{equation}

The asset owner submits $k_j$ and $s(h(F_j), k_j)$ to the Porter specified in $F_{j-1}$ via the Client in exchange for a \textit{Proof of Inclusion} $p(C_{P,i+n},k_j,F_j)$, where $n$ represents the number of roots of $C_P$ that have been sequentially produced by Porter $P$ since the creation of $C_{P,i}$. A proof of inclusion demonstrates that a key-value pair $(k_j,F_j)$ has been inserted into an associative array with root $C_{P,i+n}$. If applicable, the Porter also provides a series of \textit{Proofs of Exclusion}, $p(C_{P,i+m},k_j,\varnothing)$ for any blocks completed at time $i+m$, for $0 \leq m < n$. 

\textit{The Proof of Provenance} (POP)\footnote{The POP is an asset cryptographic audit trail, composed of the Merkle proofs for each update, demonstrating that every state transition was registered by the designated Porter and anchored in the corresponding Validator ledger roots.} $\pi_j$ for the asset comprises its POP prior to its previous update $\pi_{j-1}$ plus the Proof of Inclusion for the latest update and any Proofs of Exclusion. The POP of an asset that has not yet been transacted is empty, a \textit{null proof}.

Similarly, the Validator provides a Proof of Inclusion $p(R_{L,t},k_P,C_{P,j})$ to the Porter when it creates a new block at time $t$, to demonstrate that the Porter's root was included in the latest Validator root.\footnote{Note that in the case of heterogenous ledgers this exact mechanism will change---we assume that the Porter's TTR would exist as a transaction or entry in a finalized block's merkle root on the downstream ledger, and that this could be communicated back to the Porter for state update.} This proof of inclusion can be passed to the client to anchor the update to the Validator's ledger (Figure \ref{fig:sark_protocol}).

\subsection{Subsystems}

Sark has three key subsystems:

\subsubsection{Client}
The Client is the interface by which a USO asset can be transacted. The USO asset can be built to accommodate and wrap an off-the-shelf existing solution, such as GNU Taler, or a custom one,\footnote{As described in the Protocol section, the assumption is that one-time public keys are used, so the obvious design would be a Hierarchical Deterministic (HD) structure.} provided that it satisfies the requirements of the Porter Client API.

The Client both submits a commitment in the form of a signed hash of a new asset transaction to the corresponding Porter (specified by the asset's initial, or previous update, as described in Section 3.2) and receives the Porter's acknowledgment of the commitment, and a Proof of Inclusion that is used to extend its Proof of Provenance (POP).

\subsubsection{Porter}
Each asset is bound to a specific relay (Porter) for its next update, which is responsible for verifying its integrity, recording a commitment of the update, and extending the asset's POP. 

\begin{figure}[t!]
    \centering
    \includegraphics[
    height=5cm,
    keepaspectratio,
    ]{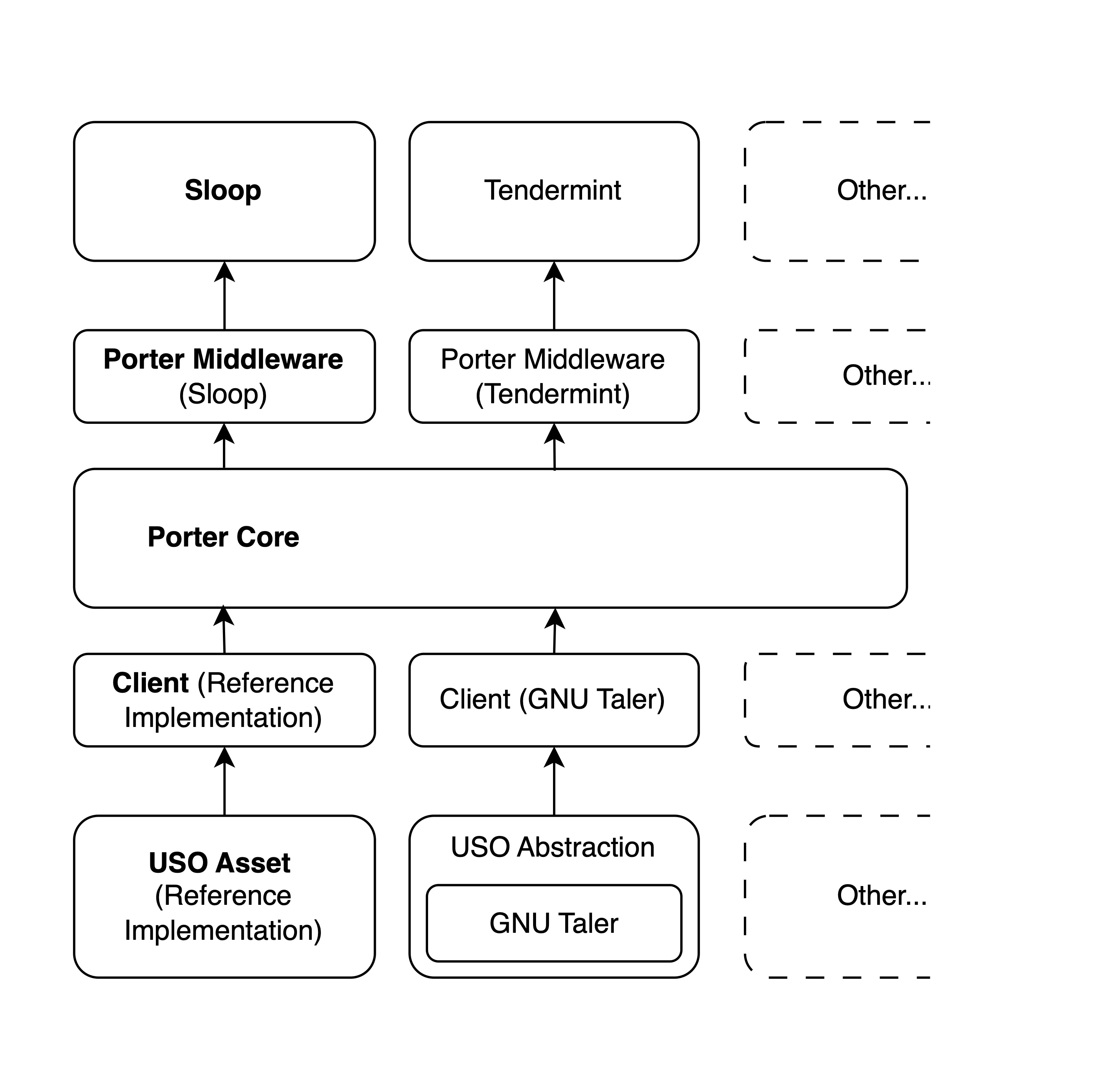}
    \caption{Sark Vertical Implementation}
    \label{fig:sark_vert}
\end{figure}

Porters operate off-chain and do not interact directly with each other.\footnote{Like an L2 client that does not participate in a consensus process.} We envisage that each Porter is paired with one (or more) validators on an underlying ledger. This can be a custom ledger for a specific use-case,\footnote{For example, Sloop in our reference architecture, or any of the private institutional blockchains currently in use.} though it is also possible to submit these aggregations of commitments to an existing blockchain (see Figure \ref{fig:sark_vert}).

Porters maintain a local ledger. They aggregate all validated transaction records within a time interval in a Merkle trie (their \textit{transaction trie}) and periodically generate Merkle roots (\textit{transaction trie roots}, or TTRs, which might be snapshots, for performance reasons---they can be indexed by the Porter for retrieval, just as ledger block roots are). Periodically, the latest Merkle root hash (TTR) is submitted to the Porter's associated Validator (or responsible blockchain) to achieve strong finality and permanent inclusion. This TTR captures the state commitments of all transactions the Porter processed during the time period, and is then included in the ledger's root hash for that block. A USO asset can only be updated with reference to its assigned Porter, preventing conflicting or duplicate state changes. This ensures protection against double-spending; a user can only include one update per one-time key, per TTR, simplifying state verification by not requiring global coordination.

\subsubsection{Validator}
Sloop is a minimal, unopinionated ledger based on the Raft protocol, whose role is to timestamp and immutably anchor each Porter's output. It does not track individual assets, transactions, or balances but simply aggregates the Merkle roots submitted by Porters, just as Porters store and aggregate the commitments provided by Clients. These roots serve as immutable commitments to each Porter's transaction trie, guaranteeing non-equivocation. 

At the end of each time period, each Sloop validator shares the Merkle root it received from its associated Porter with the Sloop network. It creates a Merkle trie by combining the entry that it received from its Porter with the entries that it received from its peers during the period, and when the root of this Merkle trie matches the Raft proposal, it signs the proposal with its signing key. Sloop broadcasts the mutually signed Merkle root via the Raft leader to all nodes. If a majority of validators agree on the root hash, then all validators append the root to their ledger.

Finally, it emits a `Commit Block' message, which can be consumed by a Porter's middleware. This message contains the ledger root hash and a proof of inclusion of the Porter's Merkle trie root.


\subsection{Implementation}


Within Sark, the Sloop blockchain is a minimal implementation that meets the Sark requirements. Sloop can be replaced with other consensus implementations, such as Tendermint, while the USO asset can be implemented as needed (as long as it retains the minimum required traits) or integrated with existing solutions (for example GNU Taler with a USO wrapper). This versatility is possible due to the modular design of the three core components of Sark: Porter Core, Porter Middleware, and the Client (see Figure \ref{fig:sark_vert}).

\begin{itemize}
    \item \textbf{Clients} allow Users to interact with USO assets that they hold. These may interact with Porters and optionally public query interfaces for a given ledger.
    \item \textbf{Porter Core} is a library for the internal operations of Porter, including receiving and aggregating transaction hashes, and sending them to validators.
    \item \textbf{Porter Middleware} provides the necessary interfaces for Porter Core to interact with validators. Porter Middleware for the Sloop case is relatively simple. However, complex implementations are possible, for example one that incorporates a Tower ABCI implementation~\cite{tower_abci} to ensure compatibility between Porter Core and Tendermint, or a middleware for Malachite.~\cite{malachite}
    \item \textbf{Sloop.} In the default case, it is envisaged that Sark would be run as a consortium system, using a permissioned ledger such as Sloop. Sloop does not store a state tree, instead concatenating hashes together to form a chain of root hashes. Since Sark has no account-based identity construct, additional data beyond block metadata is not needed. This results in a performant ledger implementation with a low overhead in terms of storage requirements, which grow linearly with time as hashes signed by the validator set are added.
\end{itemize}

A Porter, then, is a concrete implementation containing the Porter Core protocol, lifecycle hooks and a server implementation, implemented as custom middleware between Porter Core and any downstream ledgers.

Since Porters are independent, the nodes scale independently. How they choose to externalise their commitments (that is, whether they write to a permissioned ledger such as Sloop, or a public, permissionless ledger such as a Tendermint ledger, or even Ethereum or Bitcoin) is likely to be the greater constraint on time-to-finality. Comparative analysis of trie performance can be found in Appendix 1.

\subsection{Sloop Finality}

Block commitment in Sloop is implemented as a two-stage process. It uses three timeouts---the Raft heartbeat, \(T_h\), shared by all nodes; a Block timeout, \(T_b\), after which a Leader gathers commitments from Followers; and Block Commitment timeout \(T_c\), after which the Leader commits. \(T_b\) and \(T_c\) are managed by the Leader.

\begin{table}[t!]
{\renewcommand{\arraystretch}{1.5}
\quad{\setlength{\tabcolsep}{1em}
    \centering
    \caption{Sloop Node States}
    \label{tab:node_states}
    \begin{adjustbox}{width={\columnwidth},totalheight={\textheight},keepaspectratio}
    \begin{tabular}{|p{0.2\linewidth}|p{0.8\linewidth}|}\hline
          \textbf{Node State}&  \textbf{Description}\\\hline
          \raggedright \textbf{Block Creation}& Before \(T_b\). After a node's local log has been updated, it transitions into the Block Creation state. Additionally, the failure to sign and thus commit a block, or other unexpected crash faults during the \(T_c\) period could hypothetically result in no commitment, followed by a return to Block Creation.\\
          \hline
          \raggedright \textbf{Block Commitment}& After \(T_b\), but before \(T_b + \epsilon\). When followers receive a message from the Leader requesting commitments, the nodes transition into this state. Note that they will not update their local log until they receive an ack from the Leader.\\
          \hline
    \end{tabular}
    \end{adjustbox}
    }}
\end{table}
The process is simpler than Tendermint, being CFT rather than BFT, so involves only a single state transition (Table \ref{tab:node_states}), as the rest of the process, including resolving for example, Leader failure, is handled by Raft (which has a timeout in the form of the heartbeat). In Tendermint every node operates using multi-stage timeouts; in Sloop, nodes are aware of the Raft heartbeat timeout (\(T_h\)), but not the Block (\(T_b\)) or Block Commitment (\(T_c\)) timeouts. Those are managed only by the Leader. State shifts on followers occur only after receiving messages from the Leader.

Ignoring network latency as exogenous, the best case for finality is:
\[T_b + \epsilon\]

Where \(T_b\) is the configured Block timeout. \(\epsilon\) is defined as \(\epsilon \leq T_c\), where \(T_c\) is the configured timeout for the Block Commitment stage.

The upshot is that Leader failure, even in the second state, is handled gracefully. Elections must occur in a timely fashion in order for the system to be operational (a reasonable heuristic is 3-4 heartbeats), thus Leader failure at this point does not result in a huge impact on finality. Raft times the leader out, conducts an election, and, if it is in the Block Commitment state, simply re-broadcasts a request for commitments to all its new followers. In the case that a Leader fails in the \(T_b\) period, they can simply check a delta from the timestamp of the previous block commitment.\footnote{Unless the Leader is Byzantine, and changes its system clock.}

This means that the pathological case for finality is:
\[T_b + T_h + f\Delta + (f+1)\epsilon\]

Where \(f\) is the number of sequential leader failures or crash faults, and \(\Delta\) is the duration between Leader election commencing, and concluding. By design,~\cite{goodell_et_al_uso} \(T_b\), the commitment of a new top-level Merkle root, could be as long as a day.\footnote{The reasoning is that the fewer blocks created, the fewer blocks a Client would be required to check for a Proof of Inclusion, if it had no index data. In practice, it is likely that operators will desire quicker ledger finality.} Assuming the time needed for leader election can be sacrificed as part of the cost of assurance that a Porter has not equivocated, \(\Delta\) can be ignored. Elections must occur in a timely fashion in order for the system to be operational, meaning the worst-case expected in the operational system is a single Leader failure in two block commitment (\(T_c\)) phases (the \(f+1\) referenced in the prior formalization):
\[T_b + T_h + \Delta + 2\epsilon\]

\section{Analysis}

\subsection{Threat Model}

\begin{figure}[t!]
    \centering
    \includegraphics[
    width=\columnwidth,
    keepaspectratio,
    ]{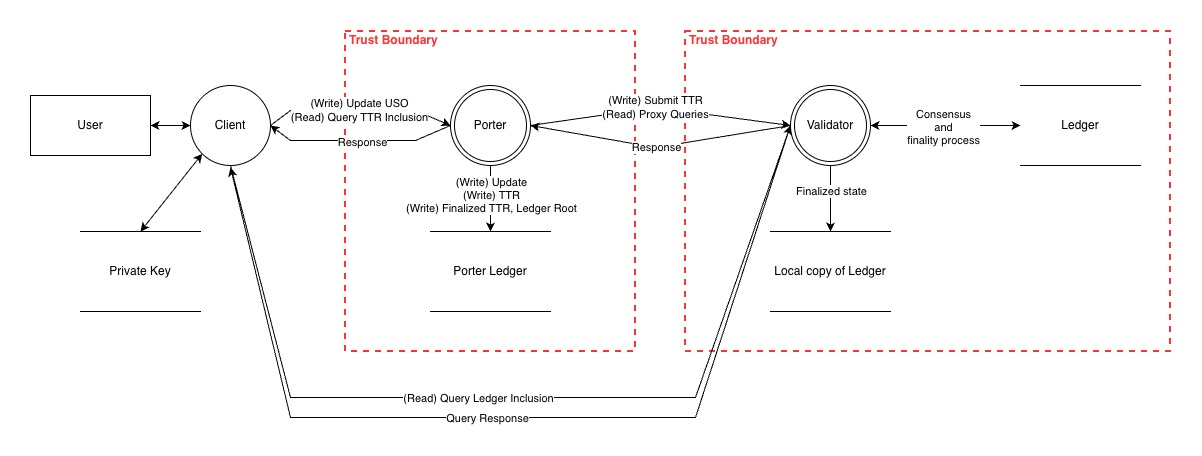}
    \caption{Sark Threat Model}
    \label{fig:sark_threat_model}
\end{figure}
Figure \ref{fig:sark_threat_model} shows the core data flows involved in managing a USO asset within the Sark system. The User is interacting through a local Client, which performs signing and blinding functions over the USO assets that are then tethered to the Porter. Table \ref{tab:stride} contextualizes this Data Flow Diagram and the USO itself within the STRIDE framework.~\cite{stride_framework} 
\begin{table}[t!]
{\renewcommand{\arraystretch}{1.5}
\quad{\setlength{\tabcolsep}{1em}
    \centering
    \caption{STRIDE Threat Analysis}
    \label{tab:stride}
    \begin{adjustbox}{width={\columnwidth},totalheight={\textheight},keepaspectratio}
    \begin{tabular}{|p{0.16\linewidth}|p{0.3\linewidth}|p{0.54\linewidth}|}\hline
          \textbf{Axis}& \textbf{STRIDE Definition}~\cite{stride_framework}& \textbf{Analysis}\\\hline
          \raggedright \textbf{Spoofing identity} & \raggedright An example of identity spoofing is illegally accessing and then using another user's authentication information, such as username and password. & The user needs to sign every update to the USO locally via their Client. The update is blinded, and handled by only the Porter specified in the USO. As the Porter is specified, the update hash will be unique for a given USO. However, the Porter itself is oblivious and just stores hashes---the data remains in the USO. The USO containing its own state in this manner is key to its Unforgeability.\\
          \hline
          \raggedright \textbf{Tampering with data} & \raggedright Data tampering involves the malicious modification of data. & All system components other than the Client operate over blinded, hashed payloads. Porters and validators sign them if necessary for audit or finality reasons, but have no knowledge of the contents.\\
          \hline
          \raggedright \textbf{Repudiation} & \raggedright Repudiation threats are associated with users who deny performing an action without other parties having any way to prove otherwise. Nonrepudiation refers to the ability of a system to counter repudiation threats. & As the state of the USO is self-contained, intuitively it should be possible to commit to more than one transfer before the new owner takes ownership of the asset (Figure \ref{fig:pop_update}). This is why the Porter must be specified by the owner before taking possession---this mechanism means either at creation, or update, only one asset history is valid, once its POP is saturated.~\cite{toda_double_spend} Commitments need to be externalized to a ledger to guard against Porter equivocation.\\ 
          \hline
          \raggedright \textbf{Information disclosure} & \raggedright Information disclosure threats involve the exposure of information to individuals who are not supposed to have access to it. & This is where obliviousness-by-design beats data protection. Unlike any scheme where data are not revealed \textit{by governance, regulation or convention}, in Sark the data are not available to subsystems due to their obliviousness.\\ 
          \hline
          \raggedright \textbf{Denial of service} & \raggedright Denial of service (DoS) attacks deny service to valid users—for example, by making a Web server temporarily unavailable or unusable. & This attack has the lowest bar for a prospective attacker, since Porters are a single point of failure for assets tethered to them, and individual Porters could equivocate, due to operator error or true malicious intent. This issue could be addressed by `communities' of Porters at the edge (Section 4.4), which could sustain read and write liveness in adversarial conditions, in the event of downtime, operator error or DoS attack, at the cost of re-introducing limited coordination.\\
          \hline
          \raggedright \textbf{Elevation of privilege} & \raggedright In this type of threat, an unprivileged user gains privileged access and thereby has sufficient access to compromise or destroy the entire system. 
          & Elevation of privilege could be describing centralization---for example, a blockchain Foundation being the cornerstone of the system. In Sark, governance centralization in consortium operation is elided via contractual arrangements. In contrast to a permissionless setting, locality reduces the feasibility of privilege escalation. However, the trade-off is that Porters are a single point of failure for the assets that have been tethered to them from the point of view of integrity. \\
          \hline
    \end{tabular}
    \end{adjustbox}
    }}
\end{table}

Table \ref{tab:stride} primarily covers the Client-to-Porter trust boundary, but discussion of Denial of Service and Repudiation require analysing the Porter-to-Ledger boundary. The key takeaway should be that equivocation, Porter failure or downtime results in the inability to transact. In the case of consortium operation, it is envisaged that such issues would be managed by SLAs and contracts. As described in Section 4.4, in any permissionless deployment a `community' of Porters---probably meeting the Practical Byzantine Fault-Tolerance threshold of 4 nodes to tolerate a single failure---would be required to disperse authority beyond a single party, ensure uptime, and externalise commitments.~\footnote{Thus reducing the opportunity for equivocation even in the absence of an external ledger; for this reason, a Porter `community' at-the-edge would be a significant improvement on the minimal architecture for consortium use described here. How this `community' would organise remains a question---would it take on a formal legal identity, or organise itself more informally, for example as a DAO?} 

\subsection{Confidentiality, Availability, Integrity}

Building on the wording of data protection legislation, threats and trade-offs can be further analysed using the three axes of the so-called `CIA triad': \textit{Confidentiality}, \textit{Availability}, and \textit{Integrity}. A discussion is presented in Table \ref{tab:cia_triad}.

`Confidentiality' is a more precise term to clarify the continuum of options that exist rather than the normal binary of `public' versus `private.' In a field study, agents made a distinction between the transmission principles and privacy norms (in the Contextual Integrity sense, building on Nissenbaum~\cite{nissenbaum}) for different agent types. Specifically, respondents indicated users or stakers might be entitled to a higher degree of confidentiality than node operators.\footnote{A node operator discussing privacy argued, ``I am not too concerned about privacy. Today's privacy for validators is good, I think it means validators should not have privacy. Sometimes people push privacy too far, they think that privacy should protect everyone. I think that privacy should protect individuals, the anonymous individuals, the users. Validators are entities, they are almost corporations. You [want] more transparency about the corporations. Like, if a validator wants to sell commissions, or voting some way, I think by default they should be public.''~\cite{KOINE:LynGoo2025}} 

\begin{table}[t!]
{\renewcommand{\arraystretch}{1.5}
\quad{\setlength{\tabcolsep}{1em}
    \centering
    \caption{The CIA Triad}
    \label{tab:cia_triad}
    \begin{adjustbox}{width={\columnwidth},totalheight={\textheight},keepaspectratio}
    \begin{tabular}{|p{0.2\linewidth}|p{0.3\linewidth}|p{0.5\linewidth}|}\hline
          \textbf{Axis}& \textbf{NIST Definition}~\cite{nist_cia}& \textbf{Discussion}\\\hline
          \raggedright \textbf{Confidentiality} & Preserving authorized restrictions on information access and disclosure, including means for protecting personal privacy and proprietary information. & Visibility and access to data. Note we use the term confidentiality rather than privacy, since privacy (after Nissenbaum~\cite{nissenbaum}) is contingent on transmission principles and social norms (i.e. what we might define as a system's governance topology), rather than its affordances for confidentiality (which might more properly exist in a network's network topology, in our prior model~\cite{KOINE:LynGoo2025}).\\
          \hline
          \raggedright \textbf{Availability} & Ensuring timely and reliable access to and use of information. & Access to the ability to read data, and where the system's design permits it, write data. \\
          \hline
          \raggedright \textbf{Integrity} & Guarding against improper information modification or destruction and ensuring information non-repudiation and authenticity. & A broader term than simply `immutability'. Integrity covers both the integrity of an asset and governance that could affect its integrity. In our prior framing of Decentralization, we might argue this is an output of both the network topology (physical structure) and governance topology (power structure) of a system.~\cite{KOINE:LynGoo2025} Identifying the locus of this integrity in terms of operative subsystems is a key point of our analysis. \\
          \hline
    \end{tabular}
    \end{adjustbox}
    }}
\end{table}
We argue that Sark has a different set of assumptions than a typical public permissionless blockchain stack.\footnote{Such as Tendermint with the Cosmos SDK (or indeed, most public, permissionless networks in existence, like Bitcoin or Ethereum).} By adopting a local-first approach and zero-knowledge proofs, Confidentiality is improved, but Availability (without the stipulation of a `community' of Porters, see Section 4.4) suffers, due to reliance on individual integrity providers (Porters), rather than a network that replicates all state.

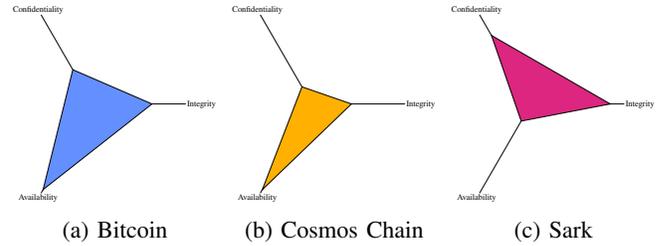
\begin{figure}[t!]
\centering
\begin{subfigure}{0.32\columnwidth}
    \resizebox{\columnwidth}{!}{%
\begin{tikzpicture}
    \coordinate (origin) at (0, 0);

    \foreach[count=\i] \radius/\dim in {4/Confidentiality,
                                        10/Availability,
                                        6/Integrity}{
        \coordinate (\i) at (\i * 360 / 3: \radius);
        \node (title) at (\i * 360 / 3: 11) {\Huge\dim};
        \draw (origin) -- (title);
    }

    \draw [fill={ibmblue}] (1)
    \foreach \i in {2,...,3}{-- (\i)} --cycle;
\end{tikzpicture}
}
    \caption{Bitcoin}
    \label{fig:bitcoin_radar}
\end{subfigure}
\hfill
\begin{subfigure}{0.32\columnwidth}
    \resizebox{\columnwidth}{!}{%
\begin{tikzpicture}
    \coordinate (origin) at (0, 0);

    \foreach[count=\i] \radius/\dim in {2/Confidentiality,
                                        10/Availability,
                                        4/Integrity}{
        \coordinate (\i) at (\i * 360 / 3: \radius);
        \node (title) at (\i * 360 / 3: 11) {\Huge\dim};
        \draw (origin) -- (title);
    }

    \draw [fill={ibmyellow}] (1)
    \foreach \i in {2,...,3}{-- (\i)} --cycle;
\end{tikzpicture}
}
    \caption{Cosmos Chain}
    \label{fig:cosmos_radar}
\end{subfigure}
\hfill
\begin{subfigure}{0.32\columnwidth}
    \resizebox{\columnwidth}{!}{%
\begin{tikzpicture}
    \coordinate (origin) at (0, 0);

    \foreach[count=\i] \radius/\dim in {8/Confidentiality,
                                        2/Availability,
                                        8/Integrity}{
        \coordinate (\i) at (\i * 360 / 3: \radius);
        \node (title) at (\i * 360 / 3: 11) {\Huge\dim};
        \draw (origin) -- (title);
    }

    \draw [fill={ibmmagenta}] (1)
    \foreach \i in {2,...,3}{-- (\i)} --cycle;
\end{tikzpicture}
}
    \caption{Sark}
    \label{fig:sark_radar}
\end{subfigure}
        
\caption{Architectural design trade-off analysis}
\label{fig:radar_plots}
\end{figure}

Integrity, as an expression of \textit{practical immutability},~\cite{lynham_goodell_immutability_springer} is improved in two key ways from many existing networks: lack of continuous reliance on global state mitigates the effect of any chain halt or rewrite,\footnote{With the obvious trade-off that finality is impacted, potentially adversely affecting Availability, unless Porters have been adapted to switch to a different ledger in times of need.} while lack of visibility of global state means the incentive to attack for hostile agents is greatly reduced. In the case of a targeted attack, identity needs to be established exogenously, and any attack conducted exogenously. While it is possible to coerce a key agent such as a Porter operator, obliviousness increases the technical barrier to such an attack; simply filtering an address is not possible in the same way.\footnote{This is a common feature of blockchains, usually achieved through filtering transactions on entry to the mempool, or a similar mechanic. A recent security report documented its use in 16 production networks,~\cite{lazarus_freeze_funds} but the number with this affordance is likely much higher.~\cite{bybit_freeze_funds}} Sark is assumed to be running in a permissioned or consortium environment in most cases.\footnote{However, it is possible, for example, to replace the Sloop Validators for Tendermint Validators, but this would involve having to incentivise them, either presumably with commercial agreements, to run via a consortium, or, in the permissionless case, some other incentive mechanism. Consider gas---which (a) introduces a marginal cost at the centre, and (b) was not discussed by validators as a factor in their business model.~\cite{KOINE:LynGoo2025}} Nevertheless, based on our definitions for Confidentiality, Availability and Integrity above, permissioned or permissionless operation does not affect the trade-off analysis that we summarize in Figure \ref{fig:radar_plots}. 

In Figure \ref{fig:radar_plots}, the ability to ex-ante identify wallets on other networks built using the default Cosmos SDK stack,\footnote{Tendermint and the Cosmos SDK, written in Go.~\cite{whatiscosmos}} including at network genesis, means that it scores worse than Bitcoin for Confidentiality. Though both rely on economic security,\footnote{That is, incentive design around rational independent agents.} the integrity guarantees of (Delegated) Proof-of-Stake are worse than Proof-of-Work.\footnote{Budish argues that Bitcoin's economic security model does not scale either, perhaps obviating the distinction.~\cite{budish10.1093/qje/qjae033}} There is no connection to entropy, and ledgers are only secured by the ex-post risk of collapse of the network's token, in the event of equivocation, attack, or rewrite.~\cite{budish2024economiclimitspermissionlessconsensus} This means that a degree of integrity has been traded for faster finality.

\subsection{Global centrality versus local centrality}

The different nature of this centrality can be expressed simply. If we take the subsystem which directly controls integrity into consideration for three systems, Bitcoin, with its miners and ledger, a Tendermint/Cosmos SDK chain utilising (Delegated) Proof-of-Stake, with its validators and ledger, and Sark, with its Porters, we can see a very different picture. Table \ref{tab:local_global_decentrality} shows the properties of each system considered in light of both the Nakamoto Coefficient~\cite{srinivasan_lee} and the Edinburgh Decentralization Index Minimum Decentralization Test (MDT).~\cite{edi}~\cite{edi_ovezik2024sokstratifiedapproachblockchain} It also identifies the subsystems in the network topology in which responsibility for integrity is concentrated; any integrity locus implies a trust locus.

\begin{table}[t!]
{\renewcommand{\arraystretch}{1.5}
\quad{\setlength{\tabcolsep}{1em}
    \centering
    \caption{Global Decentrality and Local Centrality}
    \label{tab:local_global_decentrality}
    \begin{adjustbox}{width={\columnwidth},totalheight={\textheight},keepaspectratio}
    \begin{tabular}{|p{0.25\linewidth}|p{0.25\linewidth}|p{0.25\linewidth}|p{0.25\linewidth}|}\hline
         \raggedright \textbf{Blockchain System}&  Nakamoto Consensus&  Cosmos SDK/Tendermint Chain& Sark\\\hline
         \raggedright \textbf{Example Network}&  Bitcoin&  Cosmos Hub& N/A\\\hline
         \raggedright \textbf{Subsystem Under Examination}&  Miners/Ledger&  Validators/Ledger& Porters/USO Client\\\hline
         \raggedright \textbf{Operative Threshold (Nakamoto Coefficient)}&  51\% of hash power&  33.4\% of Voting Power& 1 Porter\\\hline
         \raggedright \textbf{Attack threshold for user}& 51\% of hash power& 66.7\% of Voting Power &1 Porter\\ \hline
         \raggedright \textbf{Pathological MDT Case}& 1 Mining Pool& 1 Validator & 1 Porter\\ \hline
    \end{tabular}
    \end{adjustbox}
    }}
\end{table}
It is worth being clear about the implications of the MDT on these systems. To do this, let us consider a heuristic from Axelsen et al. for holistically analysing decentralization in DAOs: ``A DAO is only as decentralized as its crisis mode allows.''~\cite{Axelsen_2022} There is value in considering the pathological case in terms of centralization in the network topology and governance topology of networks. Since it is analysed at runtime, it accounts for the interaction of network and governance topologies (i.e. protocol and governance), and incentives (as a component of the \textit{governance topology}).

Block reversions due to the combined hash power of a single mining pool have been seen on Bitcoin.~\cite{buterinsettlementfinality} This arguably means that the pathological number of entities required to control the network is 1, failing the MDT test, since this is \textit{de facto} a single entity in charge of finality.

According to data collected from the Cosmos Ecosystem, networks can on occasion reach a single validator in control of 33.4\% of voting power, as happened on the DyDx Network shortly after launch.~\cite{cosmosnodedata} For a period of 4 months the situation was not much improved; two validators combined could stop the network. In any case, at a typical snapshot in time, a group of the biggest Cosmos chains all had a Nakamoto Coefficient of less than 10.\footnote{At a time when their on-chain activity, token price, and market cap was higher than at time of writing---see Appendix 2.} What these voting power concentration events show is that there is always a risk of takeover on these globally-connected architectures; Sark instead makes the required trust threshold to the local Porter higher, and explicit. That is, there is a substantial counterparty risk as a result of centralization that is often underappreciated, or obscured, on other ledger-based systems. Sark instead commits to the centrality of a Porter from the point-of-view of an asset holder, meaning a calculative decision about trust can be made.

A stipulation of the MDT is that it is the ultimate controlling entity that matters. Some networks balance potential centralization of voting power by generously delegating to validators and attempting to make stake weights somewhat even.\footnote{Meaning that delegations continuing or ceasing becomes a repeated game between these agents.} On paper this improves the picture, at least in terms of the Nakamoto Coefficient, which increases. However, the controlling entity is still highly central, meaning that not only is agent independence threatened, as discussed by node operators in interviews,~\cite{KOINE:LynGoo2025} but a majority of the validator set are to some degree beholden to the network's Foundation for their continued inclusion in the validator set, and indeed profitability. This means that the de facto MDT may be as low as 1 (the network's Foundation) on some networks.\footnote{In a recent high-profile network launch, Foundation delegations were used to bootstrap the voting power of most validators,~\cite{monadvision_bootstrap} before further delegations to them.~\cite{monadvision_delegations_1}~\cite{monadvision_delegations_2} This is a common practice.}

In Sark, for an Update $U$, the owner at sequence number $j$ must send the update to the Porter specified in the USO asset's last update $F_{j-1}$ (Section 3.2). Thus the locus of trust required by a user of the system shifts, from a global scope, to a local scope. We argue the focus of their interest in governance will shift as well. Although the wider Sark system---the validators and ledger---are required for finality,\footnote{Although finality arguably occurs as soon as a Porter creates an update, the prevention of equivocation requires the final write to a ledger. Note that different Porters could, in theory, write to \textit{different} ledgers.} the key point of failure is the Porter. Locality means agents must trust a Porter absolutely, as the responsibility for anchoring the validity of an update (by including it in the Porter's \textit{transaction trie}) rests with the Porter alone, and no other subsystem. Still, different agents are trusting different Porters, rather than all agents trusting a single ledger.\footnote{A rejoinder to this could be that users submit their transactions to different Validators, or RPC or API nodes. Still, the distinction remains that all Validators must replicate state exactly, while Porters only manage their own state. Thus it is a question of a user having some reliance on all nodes, versus total reliance on one node (in Sark's case, a single Porter).}

\subsection{Enhancing Availability}



\subsubsection{Communities of Porters}

Taking inspiration from DVT (Distributed Validator Technology) implementations such as Istanbul BFT that powers SSV,~\cite{moniz2020istanbulbftconsensusalgorithm} or threshold signing systems for validator nodes such as Horcrux,~\cite{horcrux} it is possible to design a minimal at-the-edge BFT consensus system for Porters to minimise the effect of crash faults, or even Byzantine faults, which we term a `community' of Porters. Using the example of a DVT-style setup, it is possible to leverage the logic of PBFT to guard against byzantine behaviour using BFT clusters of 4 Porters.\footnote{Since the $3f + 1$ stipulation of PBFT, where $f$ is the number of faulty nodes, means that 4 nodes can tolerate 1 faulty node.~\cite{pbft}} We hypothesise that such a cluster would lead to a degree of \textit{trust} in the moment, and \textit{confidence} into the future, of such an integrity source.~\cite{lynham_goodell_immutability_springer} Further, it potentially improves the authority dispersion (after Vergne~\cite{vergne}) of such an otherwise critical point of local centralization.


\subsubsection{Read-only Availability}

If data availability as a design goal is paramount,\footnote{Availability of data is key to Vergne's `decentralization.'~\cite{vergne}} then there are additional developments that can be made to the system topology. Chief among these would be an ephemeral storage mechanic employing a gossip protocol, which we describe tersely here. Porters have semi-durable storage, so it would be possible to imagine a gossip protocol between a limited community of Porter relays. Such an arrangement would mean finalized writes could be queried from any available Porter in the community by the asset owner.\footnote{We assume a data scheme that all trees are indexed under the relevant Porter; thus the owner would have the data they need to issue an API call.} Importantly, this community is free to implement its own protocol and parameters on its own terms, without economically impacting a global set of validators.


\section{Conclusion}

In this paper, we introduced Sark, a distributed blockchain system for oblivious, non-custodial management of assets with remote integrity. We described its motivation and intellectual lineage, contrasting its design and tradeoffs with existing systems that use a blockchain.

Sark makes different assumptions about decentralization and immutability, preferring local-first integrity management over global shared state. With this comes a different structure of (de)centrality, with some of the pitfalls of global centralization exchanged for a greater degree of local centralization. However, we find merit in the argument that many of the security guarantees of extant public permissionless blockchains are essentially performative,~\cite{lynham_goodell_immutability_springer}~\cite{KOINE:LynGoo2025} or at the least, in the analysis of Budish et al., ex-post in nature.~\cite{budish2024economiclimitspermissionlessconsensus}

Sark's permissioned nature largely mitigates the risk of local centrality, and we argue that its governance topology does not privilege either performative decentralization or performative trust. Instead, user trust requirements are more clearly defined and calculative: \textit{``Do I trust my Porter operator?''} Just as global state requires global governance, we hypothesise that for the average user, local state implies a greater interest in local governance (i.e. the operator of a given Porter and Validator). At a systemic level, Sark requires a lower level of trust from all agent types, due to its reduced requirement for global governance or co-ordination.

Although we have discussed Sark in the context of a known set of operators---that is, a deployment as a permissioned or consortium network---we intend to analyse its potential deployment in a `permissionless' setting in future work, for it is in this context that it can be most effectively contrasted with existing designs, and in particular their trust, confidence and institutional architectures.

\bibliographystyle{IEEEtran}
\bibliography{biblio}

@misc{toda_double_spend,
    author = {TodaQ},
    title = "{{Security Mechanisms for Protection Against Double-Spending}}",
    note = "\url{https://web.archive.org/web/20260710102659/https://engineering.todaq.net/double_spending.pdf} [Accessed: 07.07.2026]",
    addendum = "(accessed: 07.07.2026)"
}

@misc{stride_framework,
    author = {Microsoft},
    title = "{{The STRIDE Threat Model}}",
    note = "\url{https://web.archive.org/web/20260707113033/https://learn.microsoft.com/en-us/previous-versions/commerce-server/ee823878(v=cs.20)} [Accessed: 07.07.2026]",
    addendum = "(accessed: 07.07.2026)"
}

@misc{chainlink_l2,
    author = {{Chainlink Foundation}},
    title = "{{What Is Layer 2?}}",
    note = "\url{https://chain.link/education-hub/what-is-layer-2} [Accessed: 26.03.2026]",
    addendum = "(accessed: 26.03.2026)"
}

@misc{malachite,
    author = {{Informal Systems}},
    title = "{{Malachite BFT Consensus Engine}}",
    note = "\url{https://github.com/circlefin/malachite} [Accessed: 18.12.2025]",
    addendum = "(accessed: 18.12.2025)"
}

@misc{jmt_paper,
    author = {Zhenhuan Gao, Yuxuan Hu, Qinfan Wu},
    title = "{{Jellyfish Merkle Tree}}",
    note = "\url{https://web.archive.org/web/20251202144723/https://diem-developers-components.netlify.app/papers/jellyfish-merkle-tree/2021-01-14.pdf} [Accessed: 28.11.2025]",
    addendum = "(accessed: 28.11.2025)"
}

@misc{tower_abci,
    author = {{Penumbra Labs}},
    title = "{{Tower ABCI - GitHub}}",
    note = "\url{https://github.com/penumbra-zone/tower-abci} [Accessed: 28.11.2025]",
    addendum = "(accessed: 28.11.2025)"
}

@misc{horcrux,
    author = {{Strangelove Ventures}},
    title = "{{Horcrux - GitHub}}",
    note = "\url{https://github.com/strangelove-ventures/horcrux} [Accessed: 28.11.2025]",
    addendum = "(accessed: 28.11.2025)"
}

@misc{bybit_freeze_funds,
    author = {Helen Partz},
    title = "{{Bybit finds 16 blockchains with power to freeze user funds}}",
    note = "\url{https://cointelegraph.com/news/bybit-analysts-16-blockchains-freeze-user-funds} [Accessed: 28.11.2025]",
    addendum = "(accessed: 28.11.2025)"
}

@misc{lazarus_freeze_funds,
    author = {{Bybit Lazarus Lab}},
    title = "{{Blockchain Freezing Exposed: Examine the Impact of Fund Freezing Ability in Blockchain}}",
    note = "\url{https://web.archive.org/web/20251112200819/https://assets.contentstack.io/v3/assets/bltffdbacf2f22e15fa/bltda1597363a4f2a2b/69144b86424c333a34bc9fa8/2509-T68340_Security_Report_1111.pdf} [Accessed: 28.11.2025]",
    addendum = "(accessed: 28.11.2025)"
}

@misc{penumbra_docs,
    author = {{Penumbra Foundation}},
    title = "{{Penumbra Guides}}",
    note = "\url{https://guide.penumbra.zone/} [Accessed: 28.11.2025]",
    addendum = "(accessed: 28.11.2025)"
}

@misc{monadvision_bootstrap,
    author = {{Monadvision Block Explorer}},
    title = "{{Account 0x1f131Cd4066e9D79153a08cd1D311B68eb9bc602}}",
    note = "\url{https://monadvision.com/address/0x1f131Cd4066e9D79153a08cd1D311B68eb9bc602?type=Transactions} [Accessed: 28.11.2025]",
    addendum = "(accessed: 28.11.2025)"
}

@misc{monadvision_delegations_1,
    author = {{Monadvision Block Explorer}},
    title = "{{Account 0x6810126A16826718fA52DEaDA7eb979335405406}}",
    note = "\url{https://monadvision.com/address/0x6810126A16826718fA52DEaDA7eb979335405406} [Accessed: 28.11.2025]",
    addendum = "(accessed: 28.11.2025)"
}

@misc{monadvision_delegations_2,
    author = {{Monadvision Block Explorer}},
    title = "{{Account 0x5e5cD561c772968D739824AD834aECcE78e878b4}}",
    note = "\url{https://monadvision.com/address/0x5e5cD561c772968D739824AD834aECcE78e878b4} [Accessed: 28.11.2025]",
    addendum = "(accessed: 28.11.2025)"
}

@misc{nist_cia,
    author = {{National Cybersecurity Center of Excellence (NIST)}},
    title = "{{Data Integrity: Detecting and Responding to Ransomware and Other Destructive Events}}",
    note = "\url{https://www.nccoe.nist.gov/publication/1800-26/VolA/index.html} [Accessed: 28.11.2025]",
    addendum = "(accessed: 28.11.2025)"
}

@misc{edi_ovezik2024sokstratifiedapproachblockchain,
      title={SoK: A Stratified Approach to Blockchain Decentralization}, 
      author={Christina Ovezik and Dimitris Karakostas and Aggelos Kiayias},
      year={2024},
      eprint={2211.01291},
      archivePrefix={arXiv},
      primaryClass={cs.CR},
      url={https://arxiv.org/abs/2211.01291}, 
}

@misc{edi,
    author = {{University of Edinburgh}},
    title = "{{Edinburgh Decentralisation Index}}",
    note = "\url{https://blockchainlab.inf.ed.ac.uk/edi-dashboard/} [Accessed: 14.05.2025]",
    addendum = "(accessed: 14.05.2025)"
}

@misc{cosmosnodedata,
    author = {Alexander Lynham},
    title = "{{Node data for networks in the Cosmos Ecosystem}}",
    note = "\url{https://github.com/the-frey/cosmos_data/} [Accessed: 25.02.2025]",
    addendum = "(accessed: 25.02.2025)"
}

@misc{isoblockchain,
    author = "ISO",
    title = "{{Blockchain and distributed ledger technologies — Vocabulary}}",
    note  = "\url{https://www.iso.org/obp/ui/en/#iso:std:iso:22739:ed-2:v1:en} [Accessed: 01.09.2024]",
    addendum = "(accessed: 01.09.2024)"
}

@misc{buterinsettlementfinality,
    author = "Buterin, V.",
    title = "{{On Settlement Finality}}",
    note = "\url{https://blog.ethereum.org/2016/05/09/on-settlement-finality} [Accessed: 01.09.2024]",
    addendum = "(accessed: 01.09.2024)"
}

@misc{cavoukianprivacy,
    author = "Cavoukian, A.",
    title = "{{Privacy By Design: The 7 Foundational Principles}}",
    note = "\url{https://privacy.ucsc.edu/resources/privacy-by-design---foundational-principles.pdf} [Accessed: 01.09.2024]",
    addendum = "(accessed: 01.09.2024)"
}

@misc{whatiscosmos,
    author = {{Interchain Foundation}},
    title = "{{{{What is Cosmos?}}}}",
    note  = "\url{https://web.archive.org/web/20250125090007/https://v1.cosmos.network/intro} [Accessed: 25.01.2025]",
    addendum = "(accessed: 25.01.2025)"
}

@misc{ethereumnodesandclients,
    author = {{Ethereum Foundation}},
    title = "{{Nodes and Clients}}",
    note = "\url{https://ethereum.org/en/developers/docs/nodes-and-clients/} [Accessed: 01.09.2024]",
    addendum = "(accessed: 01.09.2024)"
}

@misc{srinivasan_lee,
    author = "Srinivasan, B. \& Lee, L.",
    title = "{{Quantifying Decentralization}}",
    note = "\url{https://news.earn.com/quantifying-decentralization-e39db233c28e} [Accessed: 01.09.2024]",
    addendum = "(accessed: 01.09.2024)"
}

@article{williamson1993,
author = {Williamson, Oliver},
year = {1993},
month = {02},
pages = {453-86},
title = {Calculativeness, Trust, and Economic Organization},
volume = {36},
journal = {Journal of Law and Economics},
doi = {10.1086/467284}
}

@InProceedings{goodell_et_al_uso,
author="Goodell, Geoff
and Toliver, D. R.
and Nakib, Hazem Danny",
editor="Matsuo, Shin'ichiro
and Gudgeon, Lewis
and Klages-Mundt, Ariah
and Perez Hernandez, Daniel
and Werner, Sam
and Haines, Thomas
and Essex, Aleksander
and Bracciali, Andrea
and Sala, Massimiliano",
title="A Scalable Architecture for Electronic Payments",
booktitle="Financial Cryptography and Data Security. FC 2022 International Workshops",
year="2023",
publisher="Springer International Publishing",
address="Cham",
pages="645--678",
isbn="978-3-031-32415-4"
}

@InProceedings{chaum_1983,
author="Chaum, David",
editor="Chaum, David
and Rivest, Ronald L.
and Sherman, Alan T.",
title="Blind Signatures for Untraceable Payments",
booktitle="Advances in Cryptology",
year="1983",
publisher="Springer US",
address="Boston, MA",
pages="199--203",
isbn="978-1-4757-0602-4"
}

@misc{chaum2021issuecentralbankdigital,
      title={How to Issue a Central Bank Digital Currency}, 
      author={David Chaum and Christian Grothoff and Thomas Moser},
      year={2021},
      eprint={2103.00254},
      archivePrefix={arXiv},
      primaryClass={econ.GN},
      url={https://arxiv.org/abs/2103.00254}, 
}

@article{nissenbaum,
    author = {Nissenbaum, Helen},
    year = {2004},
    month = {05},
    pages = {},
    title = {{Privacy As Contextual Integrity}},
    volume = {79},
    journal = {Washington Law Review}
}

@inbook{Bashir2022,
    author="Bashir, Imran",
    title="{{Blockchain Age Protocols}}",
    bookTitle="{{Blockchain Consensus : An Introduction to Classical, Blockchain, and Quantum Consensus Protocols}}",
    year="2022",
    publisher="Apress",
    address="Berkeley, CA",
    pages="331--376",
    isbn="978-1-4842-8179-6",
    doi="10.1007/978-1-4842-8179-6_8",
    url="https://doi.org/10.1007/978-1-4842-8179-6_8"
}

@article{vergne,
    author = {Vergne, Jean-Philippe},
    year = {2024},
    month = {07},
    pages = {115-127},
    title = {{Web3 as Decentralization Theater? A Framework for Envisioning Decentralization Strategically}},
    volume = {89},
    isbn = {978-1-83549-601-5},
    journal = {Research in the Sociology of Organizations},
    doi = {10.1108/S0733-558X20240000089010}
}

@article{budish10.1093/qje/qjae033,
    author = {Budish, Eric},
    title = {{Trust at Scale: The Economic Limits of Cryptocurrencies and Blockchains}},
    journal = {The Quarterly Journal of Economics},
    volume = {140},
    number = {1},
    pages = {1-62},
    year = {2024},
    month = {10},
    issn = {0033-5533},
    doi = {10.1093/qje/qjae033},
    url = {https://doi.org/10.1093/qje/qjae033},
    eprint = {https://academic.oup.com/qje/article-pdf/140/1/1/59814311/qjae033.pdf},
}

@inproceedings{budish2024economiclimitspermissionlessconsensus,
author = {Budish, Eric and Lewis-Pye, Andrew and Roughgarden, Tim},
title = {The Economic Limits of Permissionless Consensus},
year = {2024},
isbn = {9798400707049},
publisher = {Association for Computing Machinery},
address = {New York, NY, USA},
url = {https://doi.org/10.1145/3670865.3673548},
doi = {10.1145/3670865.3673548},
booktitle = {Proceedings of the 25th ACM Conference on Economics and Computation},
pages = {704–731},
numpages = {28},
location = {New Haven, CT, USA},
series = {EC '24}
}

@inproceedings {pbft,
    author = {Miguel Castro and Barbara Liskov},
    title = {{Practical Byzantine Fault Tolerance}},
    booktitle = {{3rd Symposium on Operating Systems Design and Implementation (OSDI 99)}},
    year = {1999},
    address = {New Orleans, LA},
    url = {https://www.usenix.org/conference/osdi-99/practical-byzantine-fault-tolerance},
    publisher = {USENIX Association},
    month = feb
}

@article{DAVIDSON_DE_FILIPPI_POTTS_2018, title={Blockchains and the economic institutions of capitalism}, volume={14}, DOI={10.1017/S1744137417000200}, number={4}, journal={Journal of Institutional Economics}, author={Davidson, Sinclair and De Filippi, Primavera and Potts, Jason}, year={2018}, pages={639–658}}

@article{DEFILIPPI2020101284,
title = {Blockchain as a confidence machine: The problem of trust \& challenges of governance},
journal = {Technology in Society},
volume = {62},
pages = {101284},
year = {2020},
issn = {0160-791X},
doi = {https://doi.org/10.1016/j.techsoc.2020.101284},
url = {https://www.sciencedirect.com/science/article/pii/S0160791X20303067},
author = {Primavera {De Filippi} and Morshed Mannan and Wessel Reijers}
}

@article{Axelsen_2022,
   title={When is a DAO Decentralized?},
   ISSN={2255-9922},
   url={http://dx.doi.org/10.7250/csimq.2022-31.04},
   DOI={10.7250/csimq.2022-31.04},
   number={31},
   journal={Complex Systems Informatics and Modeling Quarterly},
   publisher={Riga Technical University},
   author={Axelsen, Henrik and Jensen, Johannes and Ross, Omri},
   year={2022},
   month=jul, pages={51–75} }

@misc{moniz2020istanbulbftconsensusalgorithm,
      title={The Istanbul BFT Consensus Algorithm}, 
      author={Henrique Moniz},
      year={2020},
      eprint={2002.03613},
      archivePrefix={arXiv},
      primaryClass={cs.DC},
      url={https://arxiv.org/abs/2002.03613}, 
}

@incollection{KOINE:LynGoo2025,
  author = {Lynham, A. and Goodell, G.},
  title = {{Decentralization: A Qualitative Survey of Node Operators}},
  booktitle = {FCiR25 Acta},
  year = {2025},
  month = {November},
  publisher = {De Cifris Press},
  series = {De Cifris Koine},
  pages = {27--41},
  volume = {8},
  url = {https://doi.org/10.69091/koine/vol-8-P06},
  DOI = {10.69091/koine/vol-8-P06},
  ISSN = {3034-9796},
  ISBN = {979-12-81863-07-1},
  note = {{Alex Lynham, 0009-0005-0488-7651; Geoff Goodell}}
}

@InProceedings{lynham_goodell_immutability_springer,
author="Lynham, Alex
and Goodell, Geoffrey",
editor="Pal, Shantanu
and Biswas, Kamanashis
and Kanhere, Salil
and Myers, Trina
and Muthukkumarasamy, Vallipuram",
title={{Defining DLT Immutability: A Qualitative Survey of Node Operators}},
booktitle="Distributed Ledger Technology",
year="2026",
publisher="Springer Nature Singapore",
address="Singapore",
pages="97--125",
isbn="978-981-95-9230-2"
}
%




\appendix
\section{}

\subsection{Performance}

\subsubsection{Merkle Tries}

\begin{figure}[h!]
    \centering
    \includegraphics[width=0.95\columnwidth]{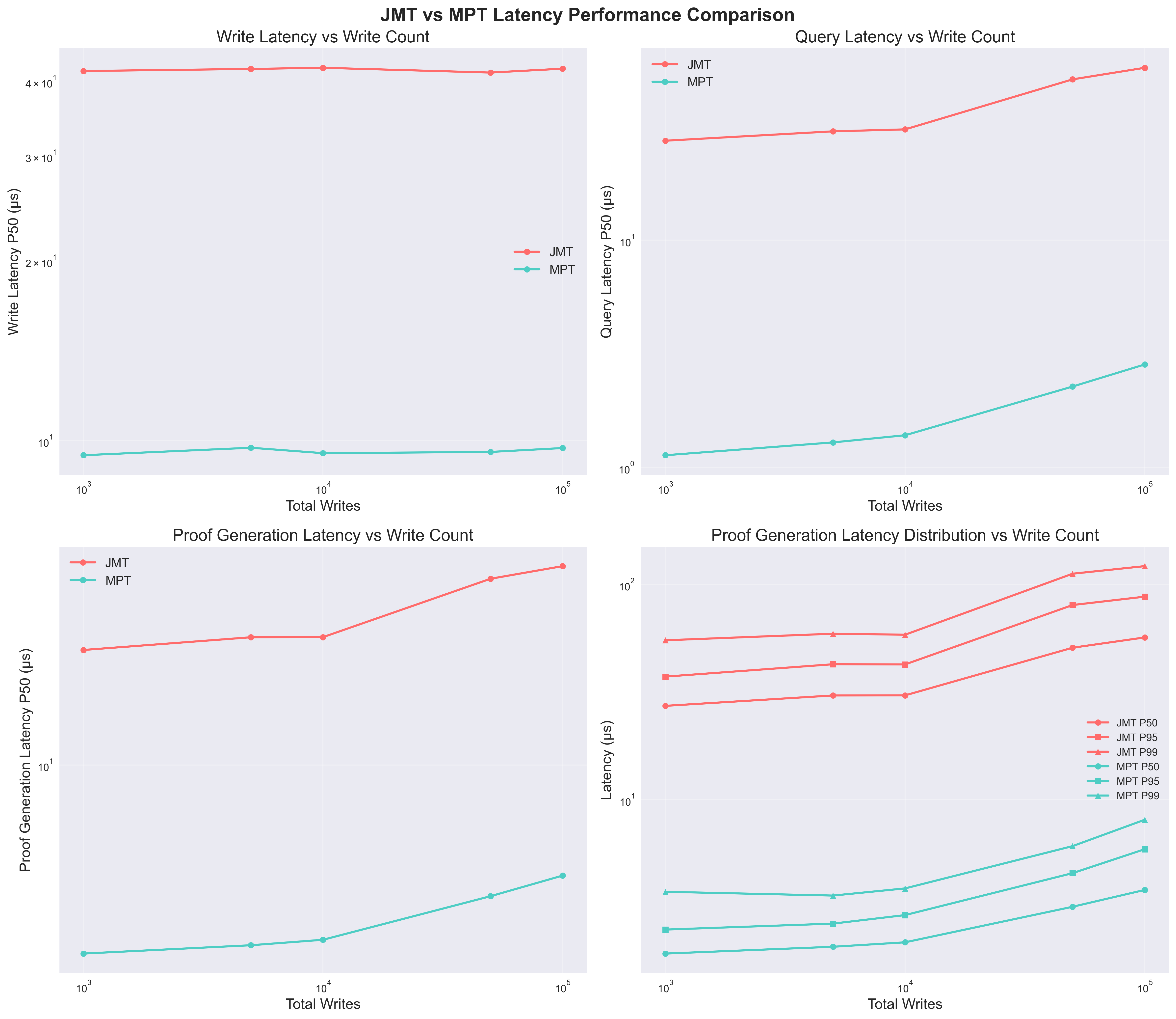}
    \caption{Latency comparison between per-block JMT and MPT}
    \label{fig:e2-latency-comparison}
\end{figure}

\begin{figure}[h!]
    \centering
    \includegraphics[width=0.95\columnwidth]{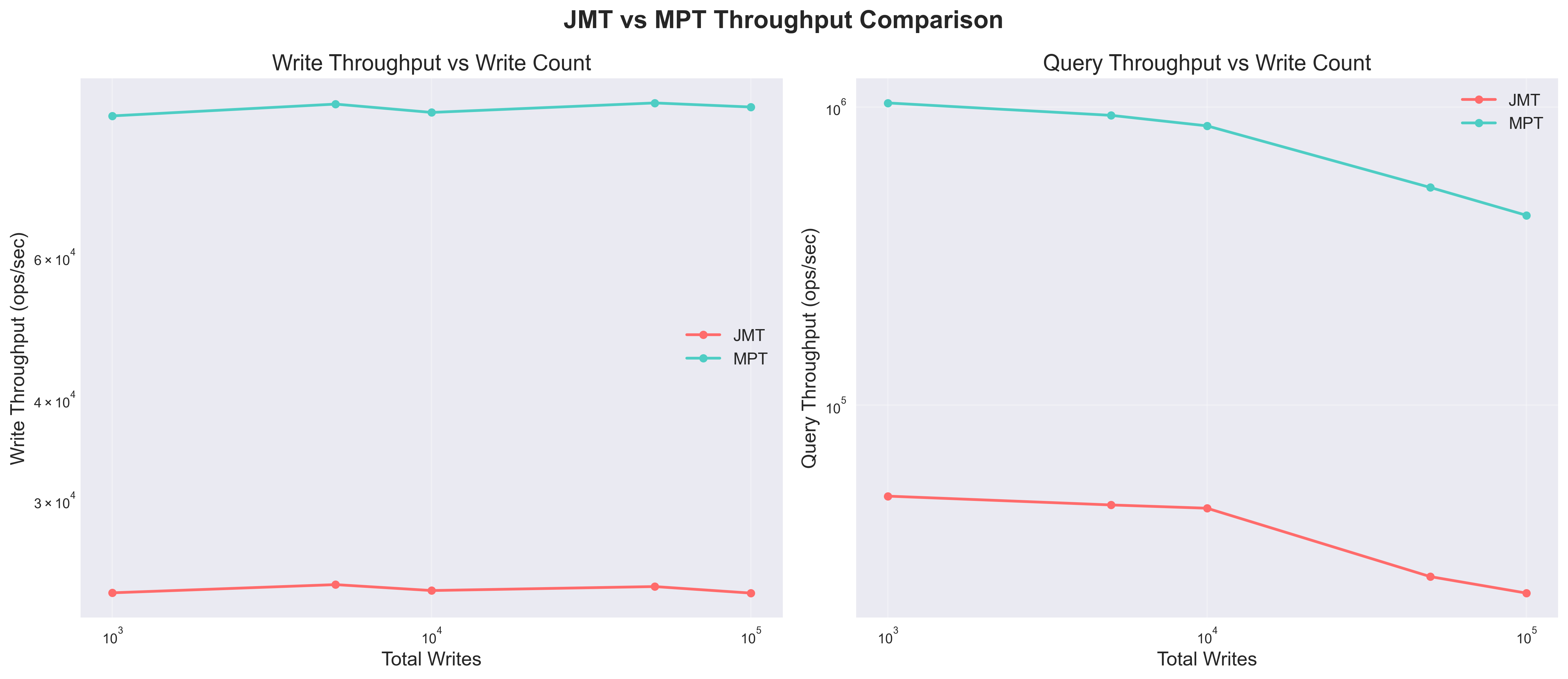}
    \caption{Throughput comparison between per-block JMT and MPT}
    \label{fig:e2-throughput-comparison}
\end{figure}

To evaluate the performance of our design, we conducted a comparison of different authenticated data structures. The comparison focusses on the distinction between per-block Jellyfish Merkle Trees~\cite{jmt_paper} and per-block Merkle Patricia Tries, enabling us to assess how the JMT's hash-centric structure contrasts with the MPT's path-compressed layout under identical block constraints.

When compared under uniformly random keys, the MPT achieves lower per-operation latency and higher throughput than the per-block JMT (Figs~\ref{fig:e2-latency-comparison} and~\ref{fig:e2-throughput-comparison}). These advantages arise from the MPT's path-compressed trie structure and its in-node value storage. By contrast, each operation in the JMT requires computing multiple cryptographic hashes: the value must first be hashed to produce a \texttt{value\_hash}, and each internal node along the path must be rehashed to incorporate the updated child commitment. Moreover, the JMT stores values out-of-tree in a separate key-value region, so each write touches both the Merkle tree nodes and the underlying value store. These factors increase CPU cost and lead to additional reads and writes within RocksDB.

Nevertheless, the JMT has structural properties that make it more suitable for systems requiring robust and interpretable state proofs. Its hash-only internal representation ensures that both membership and non-membership proofs have a uniform and bounded shape, independent of key prefixes or adversarially chosen inputs. Furthermore, the JMT natively supports copy-on-write versioning, enabling efficient historical queries and stable proof reconstruction without requiring the replay of intermediate updates. These properties align particularly well with our architecture: in Sark, the Porter layer accumulates commitments within a fixed timeout window before committing a local root, naturally producing block-scoped tries when the accumulation of these roots, that is, the chain of concatenated transaction trie roots (TTRs) is committed to a ledger (such as Sloop). 

Such windowed, snapshot-based execution of the Porter maps directly onto the JMT's versioned structure, allowing each batch to be captured as a compact, self-contained tree with predictable proof semantics. Thus, although the MPT offers superior micro-level performance under synthetic random workloads, the JMT provides a stronger and more appropriate foundation for the block-oriented, verifiable state management required by our reference design.

\subsection{Voting Power Concentration}

The voting power data used to generate these figures can be found in this public repository.~\cite{cosmosnodedata}
\begin{figure}[H]
\centering
\resizebox{\columnwidth}{!}{
\begin{tikzpicture}
\begin{axis}[%
width=\textwidth, height=4.5in,
xbar, area legend, bar width=12pt,
xmin=0, xmax=12,
symbolic y coords={
    sei,  
    osmosis, 
    juno, 
    evmos,
    dydx,
    ch, 
    akash
},
  ytick=data,
  yticklabels={Sei, Osmosis, Juno, Evmos, DyDx, Cosmos Hub, Akash},
y tick label style={align=right,text width=3cm},
xmajorgrids,
axis line style={lightgray},
major tick style={draw=none},
nodes near coords,
point meta=explicit symbolic,
node near coords style={font=\footnotesize,right=1em,pin={[pin distance=1em]180:}},
reverse legend
]

\addplot [
  fill={ibmyellow},draw=none] 
  coordinates {
    (7,sei) [7]
    (9,osmosis) [9]
    (9,juno) [9]
    (5,evmos) [5]
    (2,dydx) [2]
    (7,ch) [7]
    (6,akash) [6]
  };
\addplot [
  fill={ibmblue},draw=none] 
  coordinates {
    (7,sei) [7]
    (10,osmosis) [10]
    (9,juno) [9]
    (5,evmos) [5]
    (2,dydx) [2]
    (7,ch) [7]
    (6,akash) [6]
  };

\draw [line width=1.5pt] (current axis.south west) -- (current axis.north west);
\addlegendentry{February 2024}
\addlegendentry{January 2024}
\end{axis}
\end{tikzpicture}
}
\caption[Caption for LOF]{Nakamoto Coefficient of validators}
\label{fig:networks_snapshot}
\end{figure}

\begin{figure}[H]
\centering
\resizebox{\columnwidth}{!}{
\begin{tikzpicture}
\begin{axis}[
    width=0.95\textwidth, height=4in,
    xlabel={Months from Genesis Block},
    ylabel={Number of validators},
    xmin=0, xmax=12,
    ymin=0, ymax=28,
    xtick={0,2,4,6,8,10,12},
    ytick={0,2,4,6,8,10,12,14,16,18,20,22,24,26,28,30},
    legend pos=north west,
    ymajorgrids=true,
    grid style=dashed,
]

\addplot[
    line width=1pt,
    color=ibmblue,
    mark=square,
    ]
    coordinates {
    (0,6)
    (1,2)
    (2,2)
    (3,2)
    (4,2)
    (5,3)
    (6,4)
    (7,4)
    (8,4)
    (9,4)
    (10,4)
    (11,4)
    (12,4)
    };
    \addlegendentry{DyDx NC}

\addplot[
    line width=1pt,
    color=ibmyellow,
    mark=square,
    ]
    coordinates {
    (0,16)
    (1,4)
    (2,6)
    (3,8)
    (4,8)
    (5,10)
    (6,15)
    (7,14)
    (8,15)
    (9,15)
    (10,14)
    (11,15)
    (12,16)
    };
    \addlegendentry{DyDx TT}

\addplot[
    line width=1pt,
    color=ibmindigo,
    mark=triangle,
    ]
    coordinates {
    (0,9)
    (1,10)
    (2,11)
    (3,7)
    (4,7)
    (5,7)
    (6,8)
    (7,7)
    (8,7)
    (9,7)
    (10,7)
    (11,8)
    (12,9)
    };
    \addlegendentry{Sei NC}

\addplot[
    line width=1pt,
    color=ibmmagenta,
    mark=triangle,
    ]
    coordinates {
    (0,17)
    (1,19)
    (2,21)
    (3,21)
    (4,19)
    (5,19)
    (6,19)
    (7,19)
    (8,19)
    (9,18)
    (10,20)
    (11,21)
    (12,21)
    };
    \addlegendentry{Sei TT}

\end{axis}
\end{tikzpicture}
}
\caption{Validators required to exceed Nakamoto Coefficient (NC, 33.4\%) and Takeover Threshold (TT, 66.7\%) on the DyDx and Sei networks.}
\label{fig:thresholds}
\end{figure}

\subsection{Local Centrality}
\begin{figure}[H]
\centering
\begin{subfigure}{0.48\columnwidth}
\resizebox{\columnwidth}{!}{%
    \begin{tikzpicture}
\begin{scope}[every node/.style={circle,thick,draw}]
    \node (A) at (0,0) {V1};
    \node (B) at (0,3) {V2};
    \node (C) at (2.5,4) {V3};
    \node (D) at (2.5,1) {L1\((\eta)\)};
    \node (E) at (2.5,-1) {V4};
    \node (F) at (5,3) {V5} ;
    \node (G) at (5,1) {V6} ;
    \node (H) at (7.5,3) {Bob};
    \node (I) at (7.5,1) {Alice};
    \node (J) at (7.5,-1) {Charlie};
\end{scope}

\begin{scope}[>={Stealth[black]},
              every node/.style={fill=white,circle},
              every edge/.style={draw={ibmyellow},very thick}]
    \path [->] (A) edge node {B} (D);
    \path [->] (B) edge node {B} (D);
    \path [->] (C) edge node {B} (D);
    \path [->] (E) edge node {B} (D);
    \path [->] (F) edge node {B} (D);
    \path [->] (G) edge node {B} (D);
    \path [->] (H) edge node {tx} (F);
    \path [->] (I) edge node {tx} (G);
    \path [->] (J) edge node {tx} (G);

\end{scope}
\end{tikzpicture}
}
    \caption{Centrality of ledger in typical blockchain network (e.g. Cosmos)}
    \label{fig:blockchain_integrity_locus}
\end{subfigure}
\hfill
\begin{subfigure}{0.48\columnwidth}
\resizebox{\columnwidth}{!}{%
    \begin{tikzpicture}
\begin{scope}[every node/.style={circle,thick,draw}]
    \node (A) at (0,-1) {P1};
    \node (B) at (0,3) {P2};
    \node (C) at (2.5,4) {P3};
    \node (D) at (2.5,1) {L1\((\eta)\)};
    \node (F) at (5,3) {P4\((\eta)\)} ;
    \node (G) at (5,-1) {P5\((\eta)\)} ;
    \node (H) at (7.5,3) {Bob};
    \node (I) at (7.5,1) {Alice};
    \node (J) at (7.5,-1) {Charlie};
\end{scope}

\begin{scope}[>={Stealth[black]},
              every node/.style={fill=white,circle},
              every edge/.style={draw={ibmmagenta},very thick}]
    \path [->] (A) edge node {MR} (D);
    \path [->] (B) edge node {MR} (D);
    \path [->] (C) edge node {MR} (D);
    \path [->] (F) edge node {MR} (D);
    \path [->] (G) edge node {MR} (D);
    \path [->] (H) edge node {U} (F);
    \path [->] (I) edge node {U} (G);
    \path [->] (J) edge node {U} (G);

\end{scope}
\end{tikzpicture}
}
    \caption{Local centrality of Porters in an example implementation of a Sark system}
    \label{fig:sark_integrity_locus}
\end{subfigure}


        
\caption{The interaction of Users, Validators (V), Porters (P), and Ledgers (L), via Transactions (tx), USO updates (U, tethered to a specific Porter---the update must be sent to the Porter specified in the USO's last update) Blocks (B) and Merkle roots (MR). Integrity Locus is denoted by \(\eta\).}
\label{fig:decentrality_locality}
\end{figure}
In (\ref{fig:blockchain_integrity_locus}), all agents are dependent on the same subsystem (the Ledger) for integrity,\footnote{L1 is \(\eta(Bob, Alice, Charlie)\)} and implicitly its controlling entities. Although they submit their transactions to different Validators, responsibility for integrity (writes and durability) is global, and state is replicated; thus the locus is the Ledger, not the validators.\footnote{Perhaps, in a sense, Proof-of-Stake and Delegated Proof-of-Stake, by locking Stakers to different validators, are an attempt to create localized integrity sources at the edge, just as Porters do.} In (\ref{fig:sark_integrity_locus}), Bob, Alice and Charlie rely on different Porters for integrity,\footnote{P4 is \(\eta(Bob)\), P5 is \(\eta(Alice, Charlie)\), and L1 is \(\eta(Bob, Alice, Charlie)\)} but secondarily dependent on the same Ledger (L1) for defence against Porter equivocation. The trust requirement is greater in the local scope (i.e. if we weight \(\eta\)), and less in the global scope.\footnote{Keen readers will have inferred that there are \textit{trust} or \textit{confidence} loci, just as there are Integrity Loci in the system. The Porter is the key locus of integrity in Sark, but arguably the ledger is a key locus of \textit{trust} or \textit{confidence}, as it balances the Porter's centrality by preventing equivocation, leading to \textit{confidence} in both the calculative and (we would hypothesise) non-calculative sense.}

\end{document}